\documentclass[prb,twocolumn,showpacs,amssymb,amsmath]{revtex4}

\usepackage{graphicx}
\usepackage{bm}
\def\be{\begin{equation}}
\def\ee{\end{equation}}
\def\bea{\begin{eqnarray}}
\def\eea{\end{eqnarray}}
\newcommand{\matr}[4]{{\left(\begin{array}{cc} #1&#2\\#3&#4\\\end{array}\right)}}

\renewcommand{\vec}{\mathbf}

\renewcommand{\vr}{\vec{r}}

\newcommand{\vq}{\vec{q}}
\newcommand{\va}{\vec{a}}

\newcommand{\vX}{\vec{X}}

\newcommand{\vnabla}{\mbox{\boldmath $\nabla$}}

\newcommand{\vp}{\vec{p}}

\begin{document}
\title{Semiclassical theory of the interaction correction to the conductance of antidot arrays}

\author{Martin Schneider}
\author{Georg Schwiete}
\author{Piet W. Brouwer}
\affiliation{
Dahlem Center for Complex Quantum Systems and Institut f\"ur Theoretische Physik,
Freie Universit\"at Berlin, Arnimallee 14, 14195 Berlin, Germany
}
\date{\today}

\begin{abstract}
Electron-electron interactions are responsible for a correction to the conductance of a diffusive metal, the ``Altshuler-Aronov correction'' $\delta G_{\rm AA}$. Here we study the counterpart of this correction for a ballistic conductor, in which the electron motion is governed by chaotic classical dynamics. In the ballistic conductance, the Ehrenfest time $\tau_{\rm E}$ enters as an additional time scale that determines the magnitude of quantum interference effects. The Ehrenfest time effectively poses a short-time threshold for the trajectories contributing to the interaction correction. As a consequence, $\delta G_{\rm AA}$ becomes exponentially suppressed if the Ehrenfest time is larger than the dwell time or the inverse temperature. We discuss the explicit dependence on Ehrenfest time in quasi-one and two-dimensional antidot arrays. For strong interactions, the sign of $\delta G_{\rm AA}$ may change as a function of temperature for temperatures in the vicinity of $\hbar/\tau_{\rm E}$. 
\end{abstract}

\maketitle

\section{Introduction}
Electronic transport in weakly disordered metals is successfully described by the Boltzmann theory, in which electrons are treated as effectively classical particles moving freely between scattering events. The wave nature of electrons gives rise to a number of corrections to transport properties, such as the weak localization correction,\cite{kn:anderson1979,kn:gorkov1979} the Altshuler-Aronov interaction correction,\cite{AltshulerAronov,Altshuler85} or the universal conductance fluctuations.\cite{kn:altshuler1985b,kn:lee1985b} Weak localization results from the constructive interference of electrons propagating along time-reversed paths.\cite{ChakravartySchmid} The physical intuition behind the interaction correction is constructive interference of electron trajectories which are scattered on impurities and Friedel oscillations of the electron density.\cite{kn:rudin1997,ZalaNarozhnyAleiner} These quantum corrections become increasingly important as the temperature is lowered, the effective dimensionality 
of the sample is reduced, or as the disorder level is increased. They have a distinctive and universal dependence on external parameters, such as temperature or magnetic field, which makes them identifiable in experiments. In particular, the two quantum corrections to the conductivity, weak localization and the Altshuler-Aronov correction, can be distinguished by application of a magnetic field, since weak localization is suppressed by already a very small magnetic field, whereas the Altshuler-Aronov correction is not.

A ``classical analog'' of a disordered metal is realized in high-mobility semiconductor structures with randomly placed large antidots.\cite{kn:roukes1989,EnsslinPetroff} The absence of impurities ensures that electrons move ballistically between reflections off the antidots. The reason why these systems are referred to as classical is that the size of the antidots $a$ is much larger than the Fermi wavelength $\lambda_{\rm F}$. As a result, not only the electron's motion through the two-dimensional electron gas, but also the reflection off an antidot is described by classical mechanics. (In contrast, in a disordered metal, the size of impurities is comparable to $\lambda_{\rm F}$, so that the scattering event is strongly diffractive.) For an irregular arrangement of antidots, the classical dynamics is chaotic. Nearby trajectories separate exponentially in time, the exponential separation being characterized by the Lyapunov coefficient $\lambda$. The chaotic dynamics is essential for the existence of quantum 
corrections in this system, as it magnifies the quantum uncertainty of even a minimal wavepacket up to classical dimensions within the short time
\begin{equation}
  \label{eq:tauE}
  \tau_{\rm E} = \frac{1}{\lambda} \ln (a/\lambda_{\rm F}),
\end{equation}
thus transforming the classical dynamics into quantum-diffractive dynamics on time scales larger than $\tau_{\rm E}$.\cite{kn:aleiner1996} The time $\tau_{\rm E}$ is known as the ``Ehrenfest time''.

Since wave effects are not operative for times shorter than $\tau_{\rm E}$ --- electrons essentially move along classical trajectories up to the Ehrenfest time ---, the Ehrenfest time serves as a short-time threshold for the duration of the trajectories contributing to the quantum corrections in an antidot array. For weak localization, it was found that the correction to the conductivity is exponentially suppressed if $\tau_{\rm E}$ is larger than the dwell time $\tau_{\rm D}$, the typical time to be transmitted through the system, or the dephasing time $\tau_{\phi}$.\cite{kn:aleiner1996,Adagideli,Brouwer,kn:tian2007} In contrast, other quantum corrections, such as the universal conductance fluctuations, remain finite if $\tau_{\rm E} \gg \tau_{\rm D}$.\cite{kn:tworzydlo2004,kn:jacquod2004,kn:brouwer2006,Brouwer} 

The goal of this article is to present a theory of the Ehrenfest-time-dependence of the Altshuler-Aronov correction $\delta G_{\rm AA}$. Our analysis significantly extends a previous calculation by Kupferschmidt and one of the authors,\cite{BrouwerKupferschmidt} which studied the $\tau_{\rm E}$ dependence of the interaction correction to the conductance of a ballistic double quantum dot and found that $\delta G_{\rm AA}$ is strongly suppressed if $\tau_{\rm E}$ exceeds the dwell time $\tau_{\rm D}$ or the inverse temperature $\hbar/T$. The double quantum dot studied in Ref.\ \onlinecite{BrouwerKupferschmidt} is the simplest system with nonzero Altshuler-Aronov correction to the conductance, and is characterized by a long-range interaction, which is spatially homogeneous within each dot. The theory presented here is valid for both short-range and long-range interactions and can be applied to any geometry in which the classical electron  dynamics is chaotic --- although we will focus our discussion on the case 
of an antidot array. For the general case considered here we confirm the suppression of $\delta G_{\rm AA}$ for $\tau_{\rm E} \gg \min(\tau_{\rm D},\hbar/T)$ and we calculate the precise functional dependence of $\delta G_{\rm AA}$ on $\tau_{\rm D}$ and $T$ for finite Ehrenfest time. The explicit dependence on temperature is characteristic for the interaction correction, which has its origin in virtual processes with an energy transfer larger than temperature.

Our calculation makes use of a semiclassical formalism that starts from the saddle-point approximation around classical trajectories for the single-particle Green function. In this way, the conductance in the absence of electron-electron interactions is written as a double sum over classical trajectories that connect source and drain reservoirs.\cite{kn:jalabert1990,kn:baranger1993b} Weak localization and other quantum corrections to the conductance then follow from special configurations of trajectories, in which the two trajectories in the summation are piecewise paired, and proceed through ``crossings'' at points where the pairing is changed.\cite{RichterSieber,MuellerNJP} In the language of diagrammatic perturbation theory, segments where the trajectories are paired correspond to diffusons or cooperons, whereas the crossings correspond to Hikami boxes. The application to interacting electrons requires a modification of the formalism, which will be described in detail below.

Our analysis applies to a ``ballistic'' conductor, where the label ``ballistic'' is meant to specify that the electrons move along well-defined classical trajectories. In the literature, ``ballistic'' sometimes refers to a different limit, and several calculations of the interaction correction to the conductance have been reported for such ``ballistic limits''. Whereas the original work of Altshuler and Aronov\cite{AltshulerAronov} addressed a disordered metal with short-range scatterers in the diffusive regime $T\tau\ll 1$, the theory was generalized to account for the effects of higher temperatures $T\tau \gtrsim 1$, a regime referred to as ``ballistic''.\cite{GoldDolgopolov,DasSarmaHwang,ZalaNarozhnyAleiner} The case of a smooth disorder potential, in which scattering is predominantly forward, was considered in Ref.\ \onlinecite{GornyiMirlin}. Another type of system, where interaction corrections appear, are networks of capacitively coupled ballistic quantum dots,\cite{Beloborodov,GolubevZaikin,
KupferschmidtBrouwer} where, however, Ehrenfest-time-related phenomena can be neglected as long as $\tau_{\rm E}$ is much smaller than the dwell time in a single quantum dot. Interactions also affect the conductance through their effect on the 
weak localization correction (dephasing). Semiclassical studies of the effect 
of interaction-induced dephasing on weak localization can be found in Refs. \onlinecite{Yevtushenko,kn:tian2007,Petitjean,Whitney08}, 
 for electronic systems and in Ref.\ \onlinecite{Hartmann} for bosonic matter waves.

In Section \ref{sec:SCtheory} we present our theory of the Ehrenfest-time dependence on the interaction correction for a generic ballistic chaotic conductor. In Sec.\ \ref{sec:antidot} we then apply our formalism to an antidot array, where the classical electronic motion is diffusive on length scales much larger than the spacing between antidots, and the Coulomb interaction is dynamically screened by the diffusively moving electrons. For the antidot array, we find $\delta G_{\rm AA}\propto \exp(-\tau_{\rm E}/\tau_{\rm D}-2\pi T\tau_{\rm E}/\hbar)$ in the limit that the Ehrenfest time is larger than dwell-time and inverse temperature. For small Ehrenfest times, we recover the results of a disordered metal with quantum impurities, which show a much weaker temperature dependence (algebraic or logarithmic, depending on dimensionality). We conclude in \ref{sec:conclusion}.

\section{Semiclassical theory of the interaction correction}
\label{sec:SCtheory}
In this section we present the semiclassical description of the interaction corrections for a conductor with a well-defined chaotic classical electron dynamics. We first review the expressions for the interaction corrections to the conductance in terms of the single-particle Green function, and then apply the semiclassical approximation methods, taking into account the finite Ehrenfest time.

\subsection{Skeleton diagrams for the conductance}
\label{sec:skeleton}
\begin{figure}[t]
\includegraphics[width=2.9in]{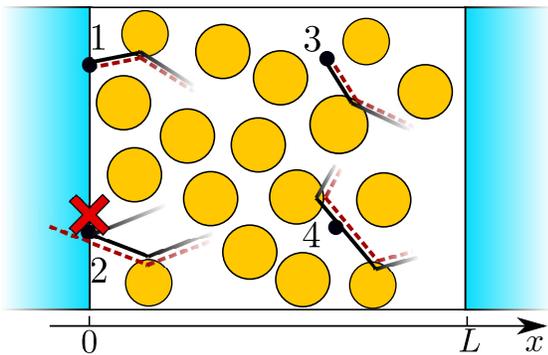}
\caption{(Color online) Schematic picture of the system under consideration: A ballistic conductor, attached to ideal leads at $x=0$ and $x=L$. In the semiclassical calculation of the conductance, one retarded and one advanced Green function are attached to a current vertex located at the interface with the leads. In a semiclassical picture, these Green functions are associated with ``retarded'' and ``advanced'' classical trajectories (solid and dashed in the figure), both of which must point into the conductor (1). A current vertex combined with two Green functions of the same kind is not possible for the calculation of the conductance: after pairing, we have trajectories that go straight into the leads (2). On the contrary, for a calculation of the conductivity, the current vertex can be anywhere inside the conductor, and pairing of retarded and advanced trajectories is possible also if two Green functions of the same kind are attached to one current vertex (3 and 4).}
\label{fig:vertex}
\end{figure}

For definiteness, we consider a two-dimensional ballistic conductor, such as a ballistic electron gas with an antidot array, in contact with reservoirs at $x=0$ and $x=L$, see Fig.\ \ref{fig:vertex}. Without interactions, we can calculate the conductance $G$ from the Kubo formula,\cite{Akkermans}
\begin{align}
 \label{eq:Kubo}
  G=&\frac{e^2 \hbar}{\pi} \int dy \int dy' \int d\xi \left(-\frac{\partial f(\xi)}{\partial\xi}\right)\nonumber\\
    &\times\left[\hat{v}_{x}{\cal G}^{\rm R}(\vr,\vr';\xi) \hat{v}_{x'} {\cal G}^{\rm A}(\vr',\vr;\xi)\right]_{\genfrac{}{}{0pt}{}{x'=0}{x=L}},
\end{align}
where $f(\xi)=1/(\exp(\xi/T)+1)$ denotes the Fermi function, 
\begin{align}
\hat{v}_x=\frac{\hbar}{2mi}\left(\overrightarrow{\partial_x}-\overleftarrow{\partial_x}\right)
\end{align}
is the velocity operator, and ${\cal G}^{\rm R}(\vr,\vr';\xi)$ and ${\cal G}^{\rm A}(\vr,\vr';\xi)$ is the retarded and advanced single-particle Green function, respectively. Retarded and advanced Green functions are related as
\begin{equation}
  {\cal G}^{\rm A}(\vr',\vr;\xi) = {\cal G}^{\rm R}(\vr,\vr';\xi)^*.
  \label{eq:GRGA}
\end{equation}

\begin{figure}[t]
\includegraphics[width=2.9in]{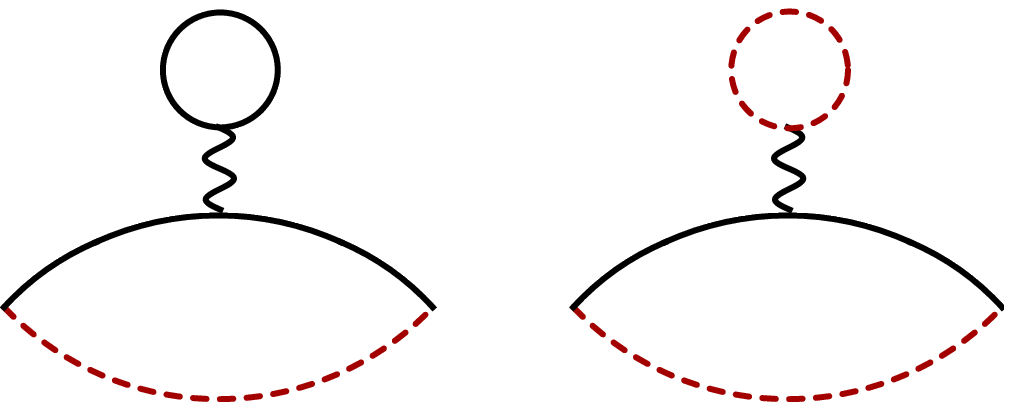}
\includegraphics[width=2.9in]{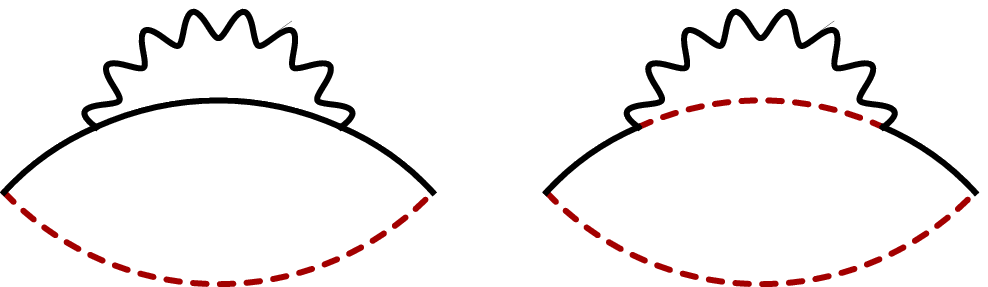}
\caption{(Color online) Hartree (upper) and Fock (lower) diagrams: solid (dashed) lines represent retarded (advanced) Green functions, the wiggly line represents the interaction. Each diagram has a counterpart with retarded and advanced Green functions interchanged.}
\label{fig:Fock}
\end{figure}

To leading (first) order in the interaction strength, the interaction correction $\delta G_{\rm AA}$ is obtained by replacing ${\cal G}^{\rm R}(\vr,\vr';\xi)$ by ${\cal G}^{\rm R}(\vr,\vr';\xi) + \delta {\cal G}^{\rm R}_{\rm F}(\vr,\vr';\xi) + \delta {\cal G}^{\rm R}_{\rm H}(\vr,\vr';\xi)$ and expanding to first order in the interaction $U$,\cite{AleinerAG,ZalaNarozhnyAleiner,BrouwerKupferschmidt} where
\begin{eqnarray}
  \label{eq:dGFock}
  \delta {\cal G}^{\rm R}_{\rm F} (\vr,\vr';\xi)&=&\int \frac{d\omega}{4\pi i}
  \int d\vr_1 d\vr_2\,
  \tanh \left(\frac{\omega-\xi}{2 T}\right) \nonumber\\
&&\mbox{} \times {\cal G}^{\rm R}(\vr,\vr_1;\xi) {\cal G}^{\rm R}(\vr_2,\vr';\xi)\nonumber\\
  &&\mbox{} \times \lbrace U^{\rm A}(\vr_1,\vr_2;\omega) {\cal G}^{\rm R}(\vr_1,\vr_2;\xi-\omega)\nonumber\\
  &&\ \ \mbox{} -U^{\rm R}(\vr_1,\vr_2;\omega) {\cal G}^{\rm A}(\vr_1,\vr_2;\xi-\omega)\rbrace, \nonumber \\ 
\end{eqnarray}
\begin{eqnarray}
 \label{eq:dGHartree}
  \delta {\cal G}^{\rm R}_{\rm H} (\vr,\vr';\xi)&=&-2
  \int \frac{d\omega}{4\pi i} \int d\vr_1 d\vr_2 \tanh \left(\frac{\omega-\xi}{2 T}\right) \nonumber\\
   &&\mbox{} \times {\cal G}^{\rm R}(\vr,\vr_1;\xi) {\cal G}^{\rm R}(\vr_1,\vr';\xi)\nonumber\\
  &&\mbox{} \times \lbrace U^{\rm A}(\vr_1,\vr_2;0) {\cal G}^{\rm R}(\vr_2,\vr_2;\xi-\omega)\nonumber\\
  &&\ \ \mbox{} - U^{\rm R}(\vr_1,\vr_2;0) {\cal G}^{\rm A}(\vr_2,\vr_2;\xi-\omega)\rbrace, \nonumber \\
\end{eqnarray}
and with similar expressions for the advanced functions $\delta {\cal G}^{\rm A}_{\rm F}(\vr',\vr;\xi)$ and $\delta {\cal G}^{\rm A}_{\rm H}(\vr',\vr;\xi)$. In these expressions, $U^{\rm R}(\vr_1,\vr_2;\omega)$ and $U^{\rm A}(\vr_1,\vr_2;\omega)$ are the retarded and advanced interaction kernels, respectively. The interaction is taken to be zero in the leads, for $x < 0$ and $x > L$.
Such a structure represents the change of the single-particle Green function due to scattering off Friedel oscillations of the density matrix $\rho(\vr_1,\vr_2)$ (Fock) or the density $\rho(\vr_2)$ (Hartree).\cite{ZalaNarozhnyAleiner} The resulting contributions to $\delta G_{\rm AA}$ are represented diagrammatically as in Fig.\ \ref{fig:Fock}.

We would like to emphasize that we kept here only those diagrams for which one retarded and one advanced Green function are connected to each current vertex. These are the relevant diagrams for the calculation of the conductance. This is in contrast to the calculation of the interaction correction to the conductivity $\delta \sigma_{\rm AA}$, where diagrams with two Green functions of the same kind attached to the current vertex play an important role.\cite{AltshulerAronov} (The conductivity $\sigma$ is expressed in a similar way as Eq.\ \eqref{eq:Kubo}, but contains integrals over the $x$-coordinates as well, rather than fixing them to the contacts at $x=0$ and $x=L$). Although the structure of the calculation for conductance or conductivity considerably differs, the final results for these quantities are related by a geometrical factor only. In two dimensions, for a rectangular sample, one has $G=\sigma\frac{W}{L}$, where $W$ is the width of the system.

The difference between the conductance calculation and the conductivity calculation is readily seen in the semiclassical language. In that language, Green functions are associated with classical trajectories, and only terms in which ``retarded'' and ``advanced'' trajectories are paired contribute. (For more details, see below). Since the leads are assumed to be free of disorder and without electron-electron interactions, both the retarded and the advanced trajectories at the positions $\vr$ and $\vr'$ in Eq.\ (\ref{eq:Kubo}) must point into the conductor if the conductance is calculated. On the other hand, for a current vertex in the system's interior, pairing of advanced and retarded trajectories is still possible even if two Green functions of the same kind are attached to the current vertex, see Fig.\ \ref{fig:vertex}.

The fact that different diagrams are needed for the calculation of conductance and conductivity is well known, the same is true for the Drude conductance and conductance fluctuations of a disordered metal (see Ref.\ \onlinecite{KaneSerotaLee}). For instance, for the calculation of the Drude conductivity $\sigma_0$ in a metal with short-ranged disorder there is no need to dress the diagram with an impurity ladder, while for the classical conductance $G_0$, the diagram dressed with an impurity ladder, i.e., a diffuson, is most relevant. For the Drude conductivity one might argue that the distance between the current vertices is of the order of the mean free path, since the two Green functions decay on this scale. The diffuson in turn is long-ranged, and hence needed to describe propagation from one lead to the other, as required for the conductance.

We also note that the calculation of the conductance as it is outlined here is similar to the calculation of the density-density correlation functions. \cite{Finkelstein,Castellani} A subtle point in this regard is the existence of additional corrections in the calculation of the density-density correlation function, namely vertex corrections and the so-called wave-function renormalization. In the description developed below, both of them appear to vanish. For the density-density correlation function, in turn, vertex corrections and the wave function renormalization cancel each other, so that these corrections do not lead to a net change of the result in either case.

\subsection{Semiclassical theory}

The conductance $G$ depends on the precise locations of antidots, the system boundary, and on the Fermi energy. We now employ a semiclassical analysis in order to identify those contributions to the conductance that remain after an average over the Fermi energy.

Starting point is the semiclassical expression of the Green function ${\cal G}^{\rm R}(\vr,\vr';\xi)$ as a sum over classical trajectories $\alpha$ from $\vr'$ to $\vr$ at energy $\xi$,\cite{Gutzwiller}
\begin{equation}
 \label{eq:SemiGreen}
  {\cal G}^{\rm R}(\vr,\vr';\xi)=\frac{2\pi}{(2\pi i\hbar)^{3/2}}\sum_{\alpha} A_{\alpha} e^{i \mathcal{S}_{\alpha}/\hbar}.
\end{equation}
Here, $\mathcal{S}_{\alpha}(\vr,\vr';\xi)$ is the classical action corresponding to the trajectory $\alpha$, which has the properties 
\begin{equation}
  \frac{\partial \mathcal{S}_{\alpha}}{\partial \vr}=\vp_{\alpha},\ \
  \frac{\partial \mathcal{S}_{\alpha}}{\partial \vr'}=-\vp'_{\alpha},
\end{equation}
and 
\begin{equation}
  \frac{\partial \mathcal{S}_{\alpha}}{\partial \xi}=\tau_{\alpha},
\end{equation}
where $\vp_{\alpha}$ and $\vp'_{\alpha}$ denotes the momentum at the end and beginning of $\alpha$, respectively, and $\tau_{\alpha}$ is the duration of the trajectory. 
The stability amplitude $A_{\alpha}$ is given by $A_{\alpha}=\sqrt{|\det(D_{\alpha})|}$, with
\begin{equation}
 D_{\alpha}=\matr{\frac{\partial^2\mathcal{S}_{\alpha}}{\partial \vr'\partial\vr}}{\frac{\partial^2\mathcal{S}_{\alpha}}{\partial \vr'\partial \xi}}{\frac{\partial^2\mathcal{S}_{\alpha}}{\partial \xi \partial\vr}}{\frac{\partial^2\mathcal{S}_{\alpha}}{\partial \xi^2}}.
\end{equation}
The semiclassical Green function further contains an additional phase-shift, the so-called Maslov index,\cite{Gutzwiller} which we omitted because it does not play a role in our considerations.
The semiclassical expression for advanced Green function follows from Eq.\ (\ref{eq:GRGA}).

Using the semiclassical Green function \eqref{eq:SemiGreen} we express the interaction correction $\delta G_{\rm AA}$ as a fourfold sum over classical trajectories. We refer to these trajectories as ``retarded'' or ``advanced'', depending on the type of the Green function that they originate from. The summation over classical trajectories can be simplified for a system with chaotic classical dynamics: In this case, the classical trajectory and hence the classical action depend very sensitively on the initial conditions. On the other hand, in the semiclassical limit $\hbar\rightarrow 0$ only configurations of trajectories with sum of the actions of the ``retarded'' trajectories {\em systematically} equal to the sum of the actions of the ``advanced'' trajectories up to a difference $\Delta\mathcal{S}$ of the order of $\hbar$ contribute substantially to the conductance. This occurs only if the ``retarded'' and ``advanced'' trajectories are piecewise paired, whereby they can exchange ``partners'' only at a 
``small-angle encounter'',\cite{RichterSieber} at which two pairs meet to within a phase-space distance of order $\hbar^{1/2}$. 

For the remaining summation over trajectories, we use a sum rule that expresses the summation over trajectories $\alpha$ between positions $\vr'$ and $\vr$ and at energy $\xi$ in terms of an integral over the trajectory's duration $t$, the initial and final momenta $\vp'$ and $\vp$, as well as a ``trajectory density'' $\rho_{\xi}(\vX'\to\vX;t)$ between the phase-space points $\vX'=(\vr',\vp')$ and $\vX=(\vr,\vp)$,\cite{ArgamanPRL,ArgamanPRB}
\begin{eqnarray}
 \label{eq:sumrule}
 \lefteqn{\sum_{\alpha:\vr'\rightarrow \vr;\xi} A_{\alpha}^2 f(\vp'_{\alpha},\vp_{\alpha},\tau_{\alpha})} && \\
&=& \int_{0}^{\infty} dt \int d\vp'_{\xi} \int d\vp_{\xi} 
  \rho_{\xi}(\vX'\rightarrow\vX;t) f(\vp',\vp,t),\nonumber
\end{eqnarray}
see App.\ \ref{sec:sumrule} for details.
Here $f$ is an arbitrary function. The initial point in phase space $\vX'=(\vr',\vp')$, together with the Hamilton function $H$ uniquely determines the classical trajectory, and after a time $t$ this trajectory has reached the phase space point $\vX(t)=(\vr(\vr',\vp';t),\vp(\vr',\vp',t))$. The trajectory density 
\begin{eqnarray}
  \rho(\vX'\rightarrow\vX;t) &=&
  \delta[\vX-\vX(t)]  \\ &=& \nonumber
  \delta[\vr-\vr(\vr',\vp';t)]
  \delta[\vp-\vp(\vr',\vp';t)] 
\end{eqnarray}
selects then only those phase space points which are connected by a trajectory of duration $t$. At fixed energy $\xi$, the momentum integrations are restricted to the energy shell, $d\vp_{\xi}=d\vp \delta(\xi-H(\vp,\vr))$, and for the trajectory density we may factor out the part which ensures energy conservation,  
\begin{eqnarray}
  \rho(\vX'\rightarrow\vX;t)&=&\rho_{\xi}(\vX'\rightarrow\vX;t)
  \delta[H(\vX)-H(\vX')].~~~
\end{eqnarray}
The factor $A_{\alpha}^2$ provides the Jacobian for this transformation. 

Following the procedure outlined so far, we obtain an expression in terms of trajectory densities which, strictly speaking, is a sum of $\delta$-functions. We then replace the exact trajectory density $\rho_{\xi}$ by a coarse-grained smooth density $\overline{\rho}$.\cite{kn:smilansky1992,kn:argaman1993} The coarse graining takes place with respect to small fluctuations of the initial and final phase space points and/or the positions of the scattering discs or the system's boundaries. In the regime $\lambda\tau_{\rm D}\gg 1$, where the chaotic dynamics has fully developed, the classical dynamics is essentially stochastic, which justifies the coarse graining procedure. The coarse-grained trajectory density 
\begin{equation}
  \overline{\rho_{\xi}}(\vX' \to \vX;t)
  = P(\vX,\vX';t)
\end{equation}
can be identified with the probability density $P(\vX,\vX';t)$ that a particle originating at the phase space point $\vX'=(\vr',\vp)$ is found at the phase space point $\vX=(\vr,\vp)$ after a time $t$. (Since we are interested in the regime where temperature is much smaller than Fermi energy, we drop the dependence of the classical propagators on $\xi$.) For the case of antidot arrays, this probability density is described by a diffusion equation.

The Drude conductance is obtained by keeping only pairs of classical trajectories that connect source and drain reservoirs. Following the steps described above, we find
\begin{align}
 \label{eq:DrudeSemiclassic}
 G_0=&\frac{e^2}{2\pi^2\hbar^2}\int_0^{\infty} dt \int dy dy' \int d\vp_{\xi} d\vp'_{\xi}\nonumber\\
 &\times\left[v_{x} P(\vX,\vX';t) v'_{x}\right]_{\genfrac{}{}{0pt}{}{x'=0}{x=L}}
\end{align}
We now turn to the semiclassical calculation for the interaction correction.

\subsection{Fock contribution}

\begin{figure}[t]
\includegraphics[width=2.9in]{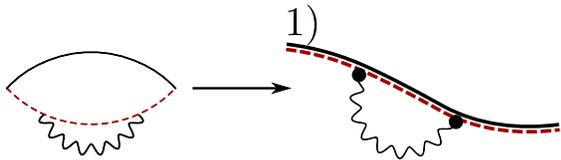}
\caption{(Color online) Configuration of trajectories relevant for the first diagram to the Fock contribution.}
\label{fig:diag1}
\end{figure}
We start with the Fock contribution to $\delta G_{\rm AA}$, which is given by the two lower diagrams in Fig.\ \ref{fig:Fock}, together with their counterparts, which are obtained by interchanging the retarded and advanced Green functions. We first consider the conductance correction $\delta G_{\rm AA}^{\rm F,1}$ from the lower left diagram and its counterpart, which reads
\begin{widetext}
\begin{eqnarray}
  \label{eq:F1}
  \delta G_{\rm AA}^{{\rm F},1} &=&-\frac{e^2 \hbar}{\pi}  \int d\xi \left(-\frac{\partial f(\xi)}{\partial\xi}\right) \int \frac{d\omega}{2\pi}\tanh \left(\frac{\omega-\xi}{2 T}\right)
  \mathrm{Im}\left\lbrace \int dy \int dy' \int d\vr_1 d\vr_2  U^{\rm R}(\vr_1,\vr_2;\omega)
  \right. \nonumber\\ && \left. \vphantom{\int} \mbox{} \times
  \left[\hat{v}_{x}{\cal G}^{\rm R}(\vr,\vr';\xi) \hat{v}_{x'}  {\cal G}^{\rm A}(\vr',\vr_2;\xi)
  {\cal G}^{\rm A}(\vr_2,\vr_1;\xi-\omega) {\cal G}^{\rm A}(\vr_1,\vr;\xi)\right]_{\genfrac{}{}{0pt}{}{x'=0}{x=L}}\right\rbrace.
\end{eqnarray}
After insertion of the semiclassical expression Eq.\ \eqref{eq:SemiGreen} for the Green functions, we obtain a sum over one retarded and three advanced trajectories. In the semiclassical limit, a convolution of Green functions is customarily calculated using the stationary phase approximation. For this one first needs to determine the configurations of trajectories which make the total action stationary. This results in a factor $\propto e^{i\mathcal{S}_{\rm st}}$, where $\mathcal{S}_{\rm st}$ is obtained by inserting the stationary configuration into the total action. Integration over quadratic fluctuations around the stationary configurations then renders the prefactor. In the present case the convolution of Green functions is accompanied by the interaction propagator. One might expect, that the interaction propagator affects the stationary trajectories, such that they no longer connect to a single classical trajectory, as in the case without interaction. However for the calculation of the conductance, we 
need to pair the advanced trajectories with the retarded one, see Fig.\ \ref{fig:diag1}. Hence, performing the integration over $\vr_1$ and $\vr_2$ in Eq.\ \eqref{eq:F1} within stationary phase approximation, we only take into account stationary configurations that connect to a single classical trajectory. The detailed calculation is carried out in Appendix \ref{sec:convolutionrule} and has the result
\begin{eqnarray}
 \label{eq:convolution}
  \lefteqn{\int d\vr_1 d\vr_2  {\cal G}^{\rm A}(\vr',\vr_2;\xi) {\cal G}^{\rm A}(\vr_2,\vr_1;\xi-\omega) {\cal G}^{\rm A}(\vr_1,\vr;\xi)
  U^{\rm R}(\vr_1,\vr_2;\omega)
  }~~~~~~~~~~~ \nonumber \\
  &=&-\frac{1}{\hbar^2} \frac{2\pi}{(-2\pi i\hbar)^{3/2}} \sum_{\alpha:\vr'\rightarrow\vr;\xi} A_{\alpha} e^{- i \mathcal{S}_{\alpha}/\hbar } 
  \int_{0}^{\tau_{\alpha}} dt \int_0^{t} dt' 
  U^{\rm R}(\vr_{\alpha}(t),\vr_{\alpha}(t');\omega) e^{i\omega(t-t')/\hbar},
\end{eqnarray}
where $\vr_{\alpha}(t)$ is the coordinate of trajectory $\alpha$ after time $t$.
The integration over time reflects the freedom to choose $\vr_1$ and $\vr_2$ anywhere along the trajectory $\alpha$; the factor $e^{i\omega(t-t')/\hbar}$ takes into account the action difference at different energies, $\mathcal{S}_{\alpha}(\xi-\omega)=\mathcal{S}_{\alpha}(\xi)-\omega\tau_{\alpha}$ (for $\omega\ll\xi$). 
With $\vX_{\alpha}(t) = (\vr_{\alpha}(t),\vp_{\alpha}(t))$ we may rewrite
\begin{align}
 \label{eq:uX1X2}
 U^{\rm R}(\vr_{\alpha}(t),\vr_{\alpha}(t');\omega)=\int d\vX_1 d\vX_2 \rho_{\xi}(\vX'_{\alpha}\rightarrow \vX_1;t)
  \rho_{\xi}(\vX'_{\alpha}\rightarrow \vX_2;t') U^{\rm R}(\vr_1,\vr_2;\omega),
\end{align}
where $d\vX=d\vr d\vp_{\xi}$ is an integration over phase-space points on the energy shell, and $\vX'_{\alpha}=\vX_{\alpha}(0)$ is the initial phase-space point of trajectory $\alpha$.

After inserting Eqs.\ \eqref{eq:convolution} and \eqref{eq:uX1X2} into Eq.\ \eqref{eq:F1}, and upon applying the semiclassical approximation to the retarded Green function as well, the interaction correction $\delta G^{\rm F,1}_{\rm AA}$ is expressed as a double sum over trajectories $\alpha$ and $\beta$ running from $\vr'$ to $\vr$. Only diagonal terms with $\alpha = \beta$ are systematically nonzero, so that we only keep these. Again making use of the sum rule (Eq.\ \eqref{eq:sumrule}) we express $\delta G_{\rm AA}^{\rm F,1}$ as an integral over the two intermediate phase space points $\vX_1$ and $\vX_2$. Before the exact trajectory densities can be replaced by their coarse-grained versions, we split the classical trajectories into uncorrelated segments using the equality
\begin{align}
 \int_{0}^{\infty} dt \int_{0}^{t} dt' \rho_{\xi}(\vX_0\rightarrow \vX_1;t) \rho_{\xi}(\vX_0\rightarrow \vX_2;t')
=\int_{0}^{\infty} dt_1 \rho_{\xi}(\vX_0\rightarrow \vX_2;t_1) \int_{0}^{\infty} dt_2 \rho_{\xi}(\vX_2 \rightarrow \vX_1;t_2).
\end{align}
After coarse-graining, the expression for $\delta G^{\rm F,1}_{\rm AA}$ involves the probability densities $P(\vX_1\to\vX_2;t)$ for the chaotic classical motion. The expression can be further simplified by introducing
 \begin{eqnarray}
   P_{\rm in}(\vX) &=&\int dy' \int d\vp'_{\xi} \int_{0}^{\infty} dt \left[ v'_{x}  P(\vr',\vp'\to\vX;t) \right]_{x'=0}, \nonumber \\
 P_{\rm out}(\vX) &=& \int dy \int d\vp_{\xi} \int_{0}^{\infty} dt \left[
  P(\vX \to \vr,\vp;t) v_x\right]_{x=L},
 \end{eqnarray}
which express the probability that a trajectory at phase space point $\vX$ entered at the left contact or exits at the right contact, respectively. Using the equality
\begin{equation}
 \int d\xi \left(-\frac{\partial f(\xi)}{\partial\xi}\right) \tanh \left(\frac{\omega-\xi}{2 T}\right)=
  \frac{\partial}{\partial \omega} \left(\omega \coth\frac{\omega}{2 T}\right)
\end{equation}
we finally obtain
\begin{align}
 \label{eq:dGAA-F1}
 \delta G_{\rm AA}^{\rm F,1}=\frac{e^2}{4\pi^3\hbar^4}  \int d\omega \frac{\partial}{\partial \omega} \left(\omega \coth\frac{\omega}{2 T}\right)
   \mathrm{Im}\left\lbrace \int d\vX_1 d\vX_2  U^{\rm R}(\vr_1,\vr_2;\omega) K^{1} (\vX_1,\vX_2;\omega)\right\rbrace,
\end{align}
where we singled out the part containing classical propagators,
\begin{align}
\label{eq:K1}
 K^{1} (\vX_1,\vX_2;\omega)
 = \int_0^{\infty} dt P_{\rm out} (\vX_1) P(\vX_1,\vX_2;t) e^{i\omega t/\hbar} P_{\rm in} (\vX_2).
\end{align}

We now consider the interaction correction $\delta G^{\rm F,2}_{\rm AA}$ that corresponds to the lower right diagram of Fig.\ \ref{fig:Fock} and its counterpart obtained by switching retarded and advanced labels,
\begin{eqnarray}
 \label{eq:F2}
 \delta G_{\rm AA}^{\rm F,2} &=&-\frac{e^2 \hbar}{\pi}  \int d\xi \left(-\frac{\partial f(\xi)}{\partial\xi}\right) \int \frac{d\omega}{2\pi}\tanh \left(\frac{\omega-\xi}{2 T}\right)
   \mathrm{Im}\left\lbrace \int dy \int dy' \int d\vr_1 d\vr_2  U^{\rm R}(\vr_1,\vr_2,\omega)\right. \nonumber\\
    &&
  \mbox{} \times \left. \vphantom{\int} \left[\hat{v}_{x}{\cal G}^{\rm R}(\vr,\vr_1;\xi) {\cal G}^{\rm A}(\vr_1,\vr_2;\xi-\omega)
  {\cal G}^{\rm R}(\vr_2,\vr';\xi) \hat{v}_{x'}  {\cal G}^{\rm A}(\vr',\vr;\xi)\right]_{\genfrac{}{}{0pt}{}{x'=0}{x=L}}\right\rbrace.
\end{eqnarray}
\end{widetext}
Insertion of the semiclassical expression for the Green functions leads to a fourfold sum over two retarded trajectories (from $\vr'$ to $\vr_2$ and from $\vr_1$ to $\vr$), and two advanced trajectories (from $\vr'$ to $\vr$ and from $\vr_1$ to $\vr_2$).

\begin{figure}[t]
\includegraphics[width=2.9in]{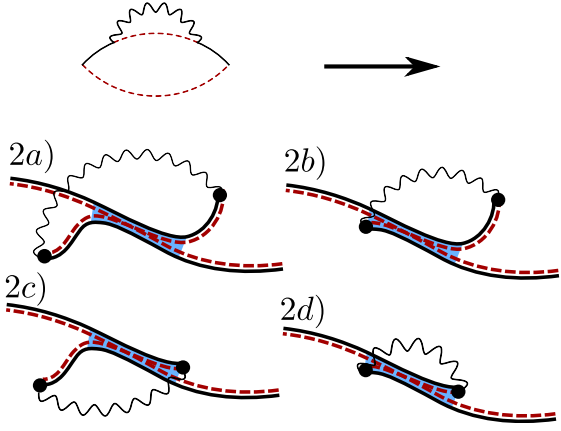}
\caption{(Color online) Configurations of trajectories relevant for the second diagram to the Fock contribution. Encounter regions are indicated in blue; here the motion of all four trajectories is correlated.}
\label{fig:diag2}
\end{figure}

Because of the specific requirements for the start and end points of the trajectories, it is not possible, to pair the trajectories one by one for their entire duration. Instead, the trajectories need to undergo a ``small-angle encounter'', in which all four trajectories are close together in phase space for at least part of their length.\cite{RichterSieber} The four possible configurations of trajectories are shown in Fig.\ \ref{fig:diag2}, where we take into account the possibilities that none, one, or both points $\vr_1$ and $\vr_2$ lie inside the encounter region. Their contributions to $\delta G_{\rm AA}$ will be denoted $\delta G_{\rm AA}^{\rm F,2a}$--$\delta G_{\rm AA}^{\rm F,2d}$, see Fig.\ \ref{fig:diag2}.

The summation over classical trajectories with a small-angle encounter follows the procedure outlined in Refs.\ \onlinecite{MuellerNJP, Brouwer}. We refer the reader to appendix \ref{sec:encrules} for details, and proceed with the results of that calculation. All four contributions to $\delta G_{\rm AA}$ have the same form as the contribution from the first diagram, see Eq.\ (\ref{eq:dGAA-F1}), but with different expressions for the function $K(\vX_1,\vX_2;\omega)$. For the contributions 2a--2d these expressions read
\begin{eqnarray}
 \label{eq:K2a-1}
 K^{2a} &=&
 -\int d\vX d\vX' P_{\rm in}(\vX) P(\vX,\vX_1;\omega)
  \nonumber\\ && \mbox{} \times 
  P(\vX_2,\vX';\omega) P_{\rm out}(\vX') 
  \nonumber\\ && \mbox{} \times 
  \frac{\partial}{\partial \tau_{\rm E}} \left[P(\vX',\vX;\tau_{\rm E}) e^{i\omega\tau_{\rm E}/\hbar}\right], \\
  \label{eq:K2b-1}
  K^{2b} &=&-\int d\vX P_{\rm in}(\vX) P(\vX,\vX_1;\omega) P_{\rm out}(\vX_2)\nonumber\\
  && \mbox{} \times P(\vX_2,\vX;\tau_{\rm E}) e^{i\omega\tau_{\rm E}/\hbar}, \\
    \label{eq:K2c}
  K^{2c} &=&-\int d\vX' P_{\rm in}(\vX_1) P(\vX_2,\vX';\omega) P_{\rm out}(\vX')\nonumber\\
  && \mbox{} \times P(\vX',\vX_1;\tau_{\rm E}) e^{i\omega\tau_{\rm E}/\hbar}, \\
 \label{eq:K2d-1}
  K^{2d} &=&
  - \int_0^{\tau_{\rm E}} dt  P_{\rm in}(\vX_1) P(\vX_2,\vX_1;t)
  \nonumber \\ && \mbox{} \times e^{i\omega t/\hbar} P_{\rm out}(\vX_2).
\end{eqnarray}
Here $P(\vX,\vX';\omega)$ is the Fourier transform of $P(\vX,\vX';t)$,
\begin{equation}
  P(\vX,\vX';\omega)=\int_{0}^{\infty} dt P(\vX,\vX';t) e^{i\omega t/\hbar}.
  \label{eq:FT}
\end{equation}
Taken together, Eqs.\ \eqref{eq:dGAA-F1}, \eqref{eq:K1}, \eqref{eq:K2a-1}, \eqref{eq:K2b-1}, \eqref{eq:K2c}, and \eqref{eq:K2d-1} determine the general result for the Fock contribution to $\delta G_{\rm AA}$ for finite Ehrenfest time, expressed in terms of classical propagators.

Let us briefly discuss the effect of the Ehrenfest time: Interestingly, the contribution $K^1$ does not involve a crossing and therefore shows no dependence on the Ehrenfest time. However, it is cancelled by the contribution $K^{2d}$, if the travel time between $\vX_1$ and $\vX_2$ is shorter than the Ehrenfest time. Thus, adding all contributions together, we indeed find, that effectively only trajectories with a duration longer than the Ehrenfest time are responsible for the interaction correction.

\subsection{Hartree contribution}

\begin{figure*}
\includegraphics[width=5.8in]{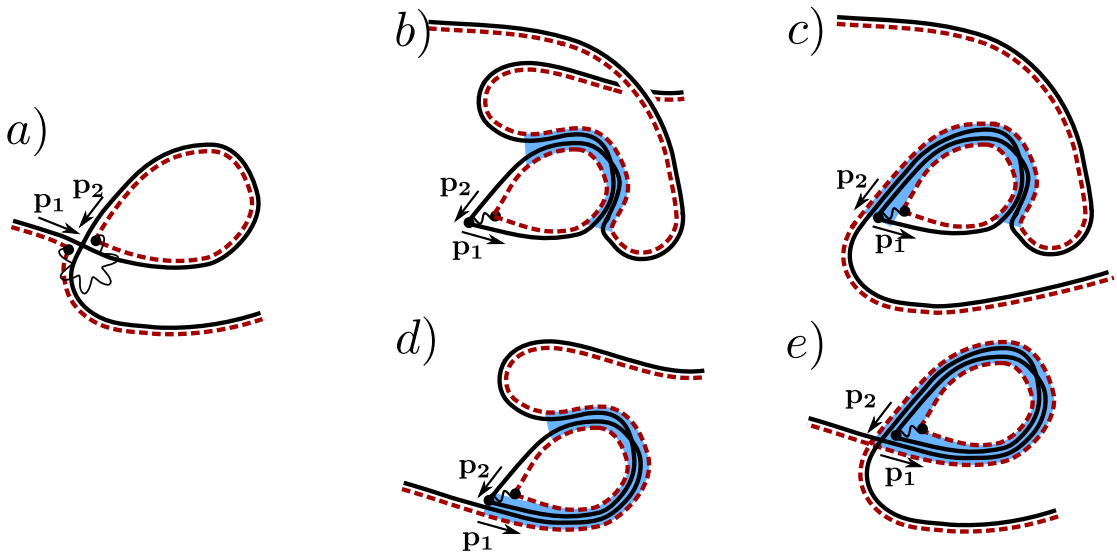}
\caption{(Color online) Configurations of classical trajectories that give the Hartree contribution to the interaction correction $\delta G_{\rm AA}$. The configurations in parts b--e contain a small-angle encounter, indicated in blue. All five configurations also contain a crossing of the classical trajectories. The momenta associated with the crossing are denoted $\vp_1$ and $\vp_2$.}
\label{fig:diag3}
\end{figure*}

The Hartree contribution to the Altshuler-Aronov correction is given by the two upper diagrams in Fig.\ \ref{fig:Fock}. Proceeding as in the case of the Fock contribution, each Green function is written as a sum over classical trajectories, which must then be piecewise paired in order to give a nonvanishing contribution to the interaction correction to the conductance. The resulting configurations of classical trajectories are shown schematically in Fig.\ \ref{fig:diag3}. The trajectory configurations of Fig.\ \ref{fig:diag3} are in one-to-one correspondence to those of Figs.\ \ref{fig:diag1} and \ref{fig:diag2} for the Fock contribution to $\delta G_{\rm AA}$: The diagram of Fig.\ \ref{fig:diag3}a corresponds to that of Fig.\ \ref{fig:diag1}, whereas the diagrams of Fig.\ \ref{fig:diag3}b--e correspond to those of Fig.\ \ref{fig:diag2}a--d.

Unlike the Fock diagrams, all diagrams for the Hartree correction involve a finite-angle crossing of the trajectories, in addition to the small-angle encounter of Figs.\ \ref{fig:diag3}b--e. Another important difference is that the action difference $\Delta {\cal S}$ for the Hartree case depends on the two positions $\vr_1$ and $\vr_2$ associated with the interaction vertex. Denoting the momenta involved in the finite-angle crossing of the trajectories by $\vp_1$ and $\vp_2$, see Fig.\ \ref{fig:diag3}, the action difference contributes an additional oscillating phase factor $e^{i(\vr_1-\vr_2)\cdot(\vp_1-\vp_2)/\hbar}$. (No fast oscillating phase factors are associated with the integration over $\vr_1$ and $\vr_2$ for the Fock diagrams.) For chaotic classical motion, the directions of the momenta $\vp_1$ and $\vp_2$ are random and uncorrelated, while the magnitude $|\vp_1| = |\vp_2| = p_{\rm F}$ is fixed by energy conservation. As a result, only a short-range component of the interaction contributes to the 
Hartree correction, and one finds the same expression for $\delta G_{\rm AA}$ as for the Fock contribution, with the replacement\cite{AltshulerAronov}
\begin{multline}
  U^{\rm R}(\vr_1-\vr_2;\omega) \to
  -2 \delta(\vr_1-\vr_2) \\   \mbox{} \times
  \left\langle \int d\vr e^{i \vr \cdot(\vp_1-\vp_2)/\hbar}
  U^{\rm R}(\vr;\omega=0) \right\rangle_{\vp_1,\vp_2},
\end{multline}
where the brackets $\langle \ldots \rangle$ indicate an average over the momenta $\vp_1$ and $\vp_2$ with $|\vp_1| = |\vp_2| = p_{\rm F}$. In case of a short-range interaction $U(\vr_1,\vr_2)\propto \delta(\vr_1-\vr_2)$, one verifies that this replacement rule leads to $\delta G_{\rm AA}^{\rm H}=-2\delta G_{\rm AA}^{\rm F}$, in agreement with Eqs.\ \eqref{eq:dGFock} and \eqref{eq:dGHartree}.

Intuitively, the interaction correction associated with the trajectory configurations of Fig.\ \ref{fig:diag3} can be interpreted as the interference of electrons that follow a classical trajectory connecting source and drain contacts, and electrons that additionally scatter from Friedel oscillations.\cite{kn:rudin1997,ZalaNarozhnyAleiner,AleinerAG} In the configuration of Fig.\ \ref{fig:diag3}a the trajectory that contains the scattering from the Friedel oscillation is shorter than that of the reference trajectory, whereas it is longer in the configurations of Figs.\ \ref{fig:diag3}b--e. The phase difference between the scattered trajectory and the reference trajectory is precisely compensated by the phase of the Friedel oscillation.\cite{AleinerAG} A similar interpretation applies to the Fock contribution, although here the scattering is from Friedel oscillations of the density matrix, not of the electron density itself.

\subsection{Coulomb interaction}

For the Coulomb interaction $U_C(\vr_1,\vr_2)=e^2/|\vr_1-\vr_2|$, due to the long-range nature, it is never sufficient to deal with the first order in perturbation theory only, and effects of dynamical screening have to be included. Hence it is not sufficient to consider only diagrams with a single bare interaction line as in Fig.\ \ref{fig:Fock}; instead one has to sum up a ladder of diagrams within the Random Phase Approximation (RPA). This analysis is explained in Refs.\ \onlinecite{Finkelstein,AltshulerAronovSSC,ZalaNarozhnyAleiner,GornyiMirlin} and it can be carried over to the semiclassical formalism without significant modifications. 

\begin{figure}[t]
\includegraphics[width=2.9in]{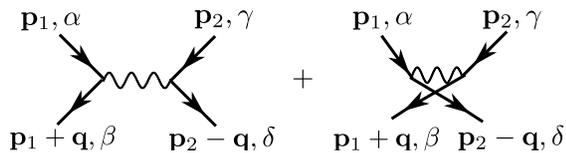}
\caption{(Color online) Lowest order scattering processes of two particles with initial momenta $\vp_1$ ($\vp_2$) and spin $\alpha$ ($\gamma$), and final momenta $\vp_1+\vq$ ($\vp_2-\vq$) and spin $\beta$ ($\delta$). Since the interaction conserves spin, the left process is proportional to $\delta_{\alpha\beta}\delta_{\gamma\delta}$ and belongs to the singlet channel, while the right process has the structure $\delta_{\alpha\delta} \delta_{\beta\gamma}=\frac{1}{2}(\delta_{\alpha\beta}\delta_{\gamma\delta}+\sum_{i}\sigma^i_{\alpha\beta}\sigma^i_{\gamma\delta})$ and therefore splits into singlet and triplet contribution.}
\label{fig:interaction0}
\end{figure}

For the purpose of including higher order interaction contributions, the separation into Hartree and Fock contributions is no longer meaningful. Instead, it is favorable to decompose the interaction into singlet and triplet channels. Hereto, we consider the interaction amplitude of a scattering process, where two particles with initial momenta $\vp_1$ ($\vp_2$) and spin $\alpha$ ($\gamma$) interact and depart with final momenta $\vp_1+\vq$ ($\vp_2-\vq$) and spin $\beta$ ($\delta$). Since in the semiclassical theory of transport only paired trajectories are relevant, we may restrict our analysis to the case $|\vq|\ll p_F$. The classification into singlet ($j=0$) and triplet part ($j=1$) then amounts to separating the interaction amplitude according to its spin structure as
\begin{equation}
 \label{eq:Uj01}
 U_{\alpha\beta\gamma\delta}=U^{(j=0)}  \delta_{\alpha\beta} \delta_{\gamma\delta} +   U^{(j=1)} \sum_{i}\sigma^i_{\alpha\beta} \sigma^i_{\gamma\delta},
\end{equation}
where $\sigma^i$ represents the Pauli-matrices ($i=x,y,z$). To lowest order, the interaction amplitude consists of the scattering processes shown in Fig. \ref{fig:interaction0}. Here, in the left process the interaction transfers a small momentum $\vq$. Such small-angle scattering appears in the Fock contribution to the interaction correction. The spin structure of this process belongs to the singlet channel. The right process allows for large-angle scattering, which appears in the Hartree contribution to the interaction correction. This process has to be split into singlet and triplet contribution, so that we end up with
\begin{equation}
 U_0^{(j=0)}(\vq)=U_C(\vq)-\frac{1}{2}\langle U_C(\vp_1-\vp_2)\rangle_{|\vp_1|=|\vp_2|=p_F},
\end{equation}
\begin{equation}
 U_0^{(j=1)}(\vq)=-\frac{1}{2}\langle U_C(\vp_1-\vp_2)\rangle_{|\vp_1|=|\vp_2|=p_F}.
\end{equation}
Here we anticipated, that for a diffusive system we may average over the directions of the momenta of the electrons.
To include screening, one then sums up the RPA-series in each channel,
\begin{equation}
 \label{eq:RPA}
 U^{(j)}(\vq,\omega)=U^{(j)}_0(\vq)- U^{(j)}_0(\vq) \Pi(\vq,\omega) U^{(j)}(\vq,\omega)
\end{equation}
where the disorder-averaged polarization operator $\Pi$ is diagonal in spin space (note, that the disorder average of the polarization operator does not involve a crossing, and therefore has no $\tau_{\rm E}$-dependence).

The discussion so far is valid for weak interaction (i.e. interaction parameter $r_s\ll1$). 
For stronger interactions, one should also include Fermi liquid effects. Then the structure of the screened interaction remains the same, but the bare interaction is now expressed in terms of Fermi-liquid parameters $F^{\rho,\sigma}_0$,
\begin{equation}
 \label{eq:UFermi}
 U_0^{(j=0)}(\vq)=U_C(\vq)+\frac{1}{2\nu}F^{\rho}_0, \quad U_0^{(j=1)}(\vq)=\frac{1}{2\nu}F^{\sigma}_0.
\end{equation}

Let's turn back to the Altshuler-Aronov correction. Applying the preceding analysis, we find that Coulomb interaction is properly included, if we calculate the Fock-type diagrams as in Fig.\ \ref{fig:Fock}, where the interaction is replaced by the effective interaction
\begin{equation}
 \label{eq:Uscreened}
 U(\vr_1,\vr_2;\omega)=U^{(j=0)}(\vr_1,\vr_2;\omega)+3 U^{(j=1)}(\vr_1,\vr_2;\omega),
\end{equation}
where the factor $3$ comes from the spin summation and accounts for the multiplicity of the triplet channel. The precise relation between  $U(\vr_1,\vr_2;\omega)$ and $U(\vq;\omega)$ follows from the solution of the diffusion equation and will be clarified in the next section (see Eq. \eqref{eq:Urq}).

\section{Interaction correction for antidot arrays}
\label{sec:antidot}
In the preceding Section, we developed the semiclassical theory of the interaction correction to the conductance, where we expressed our final results in terms of the classical propagator $P(\vX,\vX',t)$. The explicit expression of this classical propagator is determined by the geometry of the system under consideration. In semiclassical studies, two prototypical geometries are mainly investigated: ballistic quantum dots, where the classical chaotic motion results from reflections at the boundary of the dot, and antidot arrays, where the placement of artificial macroscopic scatterers leads to a chaotic dynamics. For a single ballistic quantum dot, the interaction correction vanishes. The simplest example of a geometry with a non-zero interaction correction is hence a double quantum dot, which was studied in Ref.\ \onlinecite{BrouwerKupferschmidt}. We will first show, how the results of the previous section are connected to the results of this reference. The remaining part of this section is then 
devoted to the 
interaction correction for antidot arrays, which has not been theoretically studied so far.

\subsection{Double dot}
We consider a double dot system, where two identical dots are connected by a ballistic contact of conductance $G_c$. The first (second) dot is connected to the left (right) reservoir by a ballistic contact of conductance $G_d$. The level density (i.e.\ the density of states times the dot's area) for each dot is $N$ per spin. Within each dot, the phase space is explored uniformly during the chaotic motion. Hence we might replace the integration over phase space by a sum over the dots, 
\begin{equation}
 \int d\vX =\sum_i \Omega_i,
\end{equation}
where we weigh with the available phase space volume $\Omega_1=\Omega_2=(2\pi\hbar)^2 N$ of each dot. The classical propagators are replaced by 
\begin{equation}
 P(\vX\leftarrow\vX';t)=\frac{1}{\Omega_j} P(j\leftarrow i;t).
\end{equation}
The probability $ P(j\leftarrow i;t)$ to be in dot $j$ after a time $t$, when the particle initially started in dot $i$, is then calculated from the master equation
\begin{equation}
 \label{eq:Master}
 \partial_t P(j\leftarrow i; t)=-\sum_{m} \gamma_{jm}  P(m\leftarrow i; t),
\end{equation}
where the rate matrix $\gamma$ has the form
\begin{equation}
 \gamma=\matr{\gamma_L+\gamma_{12}}{-\gamma_{12}}{-\gamma_{12}}{\gamma_R+\gamma_{12}}.
\end{equation}
Here, $\gamma_{12}=G_c/2 N e^2 $ is the rate for transitions between the dots, and $\gamma_L=\gamma_R=G_d/2 N e^2 $ is the rate for escape to the left and right lead. 
The solution to Eq.\ \eqref{eq:Master} reads
\begin{equation}
 P(j\leftarrow i; t)=\left(e^{-\gamma t}\right)_{ji}.
\end{equation}
The probability $P_{\rm in}$ ($P_{\rm out}$), that a particle in dot $i$ has entered via the left contact (leaves the system via the right contact) is given by
\begin{equation}
 P_{\rm in}(i)=\gamma_L \left(\gamma^{-1}\right)_{1i}, \quad P_{\rm out}(i)=\gamma_R \left(\gamma^{-1}\right)_{2i}.
\end{equation}
The bare interaction of the double dot system can be described by a capacitive coupling of the dots
\begin{equation}
 U_0=\frac{e^2}{C}, \quad C=\matr{C_0+C_c}{-C_c}{-C_c}{C_0+C_c},
\end{equation}
where $C_0$ describes the coupling of each dot to an external gate, and $C_c$ is the cross-capacitance between the dots. For the inclusion of screening, we make use of the polarization operator
\begin{equation}
 \Pi_{ij}(\omega)=2N\left(\delta_{ij}+\tfrac{i\omega}{\hbar}P(j\leftarrow i;\omega)\right),
\end{equation}
from which we obtain the dynamically screened interaction as
\begin{equation}
 U^{-1}(\omega)=U_0^{-1}+\Pi(\omega).
\end{equation}
For the frequencies of interest one may neglect the first term in this equation, and one obtains
\begin{equation}
 U_{ij}(\omega)=\tfrac{1}{2N}\left(1-\tfrac{i\omega}{\hbar} \gamma^{-1}\right)_{ij}.
\end{equation}
Using the expressions of this paragraph and the Eqs.\ \eqref{eq:dGAA-F1}, \eqref{eq:K1}, \eqref{eq:K2a-1}, \eqref{eq:K2b-1}, \eqref{eq:K2c}, and \eqref{eq:K2d-1} from the last section, one obtains, after some algebra, the result from Ref.\ \onlinecite{BrouwerKupferschmidt},
\begin{align}
 \delta G_{AA}=-&\frac{e^2}{2\pi\hbar} \frac{G_d G_c^2 \left(\tau_{{\rm D}-}e^{-\tau_{\rm E}/\tau_{{\rm D}+}}+\tau_{{\rm D}+}e^{-\tau_{\rm E}/\tau_{{\rm D}-}}\right)}{(G_d+2 G_c)^3}\nonumber\\
  &\times \mathrm{Im} \int \frac{d\omega}{\hbar}\frac{e^{i\omega\tau_E/\hbar} \partial_{\omega}\left(\omega \coth\frac{\omega}{2 T}\right)}{(1-i\omega\tau_{{\rm D}+}/\hbar)(1-i\omega\tau_{{\rm D}-}/\hbar)}.
\end{align}
Here, $\tau_{{\rm D}+}=2Ne^2/G_d$ and $\tau_{{\rm D}-}=2Ne^2/(G_d+2 G_c)$ are the characteristic dwell times of the double dot system (they refer to relaxtion of (anti)symmetric charge configurations). For zero Ehrenfest time, one recovers the results known from Random Matrix Theory, while for large Ehrenfest times, $\delta G_{AA}$ is suppressed as $e^{-\tau_E/\tau_{{\rm D}\pm}-2\pi\tau_E T/\hbar}$ (for more details, see Ref.\ \onlinecite{BrouwerKupferschmidt}, where the stated results are too large by a factor of two).
We note, that the Hartree contribution is zero in this case, due to the long-range nature of interaction.

\subsection{Antidot arrays}
In the following, we now apply the theory developed in Section\ \ref{sec:SCtheory} to quasi one-dimensional and two-dimensional antidot arrays. The antidot arrays consist of a ballistic electron gas with randomly placed disc-shaped scatterers of size much larger than the Fermi wavelength. The classical dynamics in such an antidot array is chaotic, and diffusive on length scales much larger than the disc size $a$ or the distance between discs. In particular, since the Ehrenfest time $\tau_{\rm E}=\lambda^{-1}\ln(a/\lambda_{\rm F})\gg \lambda^{-1}$ because of the large logarithm, and since $\lambda^{-1}$ is comparable to the transport time $\tau$, the diffusive dynamics applies for timescales down to $\tau_{\rm E}$.

A diffusively moving particle quickly loses its memory about the direction of motion, so that the classical propagators $P(\vX,\vX';t)$ depend on the positions $\vr$ and $\vr'$ associated with the phase space points $\vX$ and $\vX'$ only. This leads to significant simplifications of the general semiclassical expressions for the interaction correction $\delta G_{\rm AA}$. In order to evaluate $\delta G_{\rm AA}$ in this limit, we start by expressing the integration over momentum at fixed energy as an integration over the corresponding angle $\phi$,
\begin{equation}
 d\vp_{\xi}=(2\pi\hbar)^2 \nu \frac{d\phi}{2\pi},
\end{equation}
where $\nu$ is the density of states per spin. We then find
\begin{equation}
 P(\vX,\vX';t)=\frac{1}{(2\pi \hbar)^2 \nu} P(\vr, \vr';t),
\end{equation}
where $P(\vr \to \vr';t)$ depends on the positions only and satisfies a diffusion equation,
\begin{equation}
 \label{eq:diff}
 (\partial_{t}-D\Delta_{\vr})P(\vr, \vr';t)=\delta(t)\delta(\vr-\vr'),
\end{equation}
with diffusion coefficient $D$. For a rectangular sample of dimension $L \times W$, coupled to ideal leads at $x=0$ and $x=L$ and with insulating boundaries at $y=0$ and $y=W$, the solution of Eq.\ \eqref{eq:diff} reads
\begin{equation}
 P(\vr,\vr';t)=\theta(t)\sum_{\vq}\psi_{\vq}(\vr)\psi_{\vq}(\vr') e^{-D\vq^2 t},
\end{equation}
with the function
\begin{equation}
 \psi_{\vq} (\vr)=\sqrt{\frac{4}{LW}} \sin(q_x x) \times \begin{cases}
                                            1/\sqrt{2} & \mbox{if $q_y=0$}, \\
                                           \cos(q_y y) & \mbox{if $q_y\neq 0$}.
                                          \end{cases}
\end{equation}
The sum over $\vq$ runs over $q_x=\frac{n\pi}{L}$ with $n=1,2,...$ and $q_y=\frac{m \pi}{W}$ with $m=0,1,...$. We also use the Fourier-transformed diffusion propagator,
\begin{equation}
 \label{eq:Pomega}
 P(\vr,\vr';\omega)=\sum_{\vq}\frac{\psi_{\vq}(\vr)\psi_{\vq}(\vr')}{D \vq^2-i\omega/\hbar}.
\end{equation}
Finally, the probabilities $P_{\rm in}(\vr)$ and $P_{\rm out}(\vr)$ that a trajectory originating at position $\vr$ exits the sample on the left or on the right, respectively, are
\begin{align} 
  P_{\rm out}(\vr)= \frac{x}{L},\ \
  P_{\rm in}(\vr) = \frac{L-x}{L},
\end{align}
which may be derived from the diffusive flux 
\begin{equation}
 j_x(\vr,\vr';t)=-D\partial_x P(\vr,\vr';t),
\end{equation}
at position $\vr$ and time $t$, for a particle starting from $\vr'$ at time $t=0$. The Drude conductance (Eq. \eqref{eq:DrudeSemiclassic}) is then expressed as
\begin{equation}
 G_0=\frac{e^2}{2\pi^2\hbar^2} \int dy \int d\vp_{\xi} \left[-D\partial_{x}P_{\rm in}(x)\right]_{x=L}
\end{equation}
which gives the familiar result
\begin{equation}
 G_0=2e^2\nu D \frac{W}{L}.
\end{equation}

Let us now turn to the interaction. For the inclusion of screening effects we need the polarization operator, which, for the low frequencies at which the electron dynamics is effectively diffusive, can be expressed through the diffusion propagator,
\begin{equation}
 \Pi(\vr,\vr';\omega)=2\nu\left[\delta(\vr-\vr')+\tfrac{i\omega}{\hbar} P(\vr,\vr';\omega) \right],
\end{equation}
Using Eqs.\ \eqref{eq:RPA}, \eqref{eq:UFermi}, \eqref{eq:Uscreened}, we then find, that the effective interaction can be written as
\begin{equation}
 \label{eq:Urq}
 U^{\rm R}(\vr_1,\vr_2;\omega)=\sum_{\vq}\psi_{\vq}(\vr) \psi_{\vq}(\vr') U^{\rm R} (\vq,\omega)
\end{equation}
where $U^{\rm R}(\vq,\omega)= U^{{\rm R},(j=0)}(\vq;\omega)+3  U^{{\rm R},(j=1)}(\vq;\omega)$ is given by
\begin{equation}
 U^{{\rm R},(j=0)}(\vq;\omega)=\frac{1}{2\nu}\frac{D\vq^2-i\omega/\hbar}{D\vq^2}
\end{equation}
\begin{equation}
 U^{{\rm R},(j=1)}(\vq;\omega)=\frac{F_0^{\sigma}}{2\nu}\frac{D\vq^2-i\omega/\hbar}{D\vq^2(1+F_0^{\sigma})-i\omega/\hbar}
\end{equation}
in the singlet and triplet channel, respectively. Due to the divergence of the bare Coulomb interaction at small momenta, the interaction in the singlet channel is set by the polarization operator solely. In the triplet channel, the interaction depends on the zero angular harmonic of $F^{\sigma}$, which is the only free parameter controlling the interaction strength.

For the further calculations, it is convenient to make use of the diffusion equation, to write 
\begin{widetext}
\begin{eqnarray*}
  \frac{\partial}{\partial \tau_{\rm E}} [P(\vr,\vr';\tau_{\rm E}) e^{i\omega\tau_{\rm E}/\hbar}]
  &=& \left (D\Delta+ \frac{i\omega}{\hbar} \right)P(\vr,\vr';\tau_{\rm E})
  e^{i\omega\tau_{\rm E}/\hbar}.
\end{eqnarray*}
With the help of additional spatial integrations over delta functions, which we then replace using Eq.\ \eqref{eq:Pomega}, the interaction correction to the conductance then takes the form
\begin{eqnarray}
 \label{eq:dGAA-1}
 \delta G_{\rm AA}&=&\frac{\nu e^2}{\pi\hbar^2}  \int d\omega \partial_{\omega}\left(\omega \coth\frac{\omega}{2 T}\right)
  \mathrm{Im}\left\lbrace \int d\vr_1 d\vr_2  U^{\rm R}(\vr_1,\vr_2;\omega) K (\vr_1,\vr_2;\omega)\right\rbrace,
\end{eqnarray}
where the function $K(\vr_1,\vr_2;\omega)$ reads
\begin{eqnarray}
 \label{eq:Kdiff}
  K(\vr_1,\vr_2;\omega) &=&\int d\vr d\vr' P_{in}(\vr) P_{out}(\vr') 
  \left\{ \mathcal{D}_{\omega} P(\vr,\vr_1;\omega)\mathcal{D}'_{\omega}P(\vr_2,\vr';\omega) \int_{\tau_{\rm E}}^{\infty} dt P(\vr',\vr;t) e^{i\omega t/\hbar}
  \right. \nonumber\\ && \mbox{}
 - P(\vr,\vr_1;\omega) P(\vr_2,\vr';\omega) \mathcal{D}_{\omega} P(\vr',\vr;\tau_{\rm E}) e^{i\omega\tau_{\rm E}/\hbar}
 + \mathcal{D}_{\omega} P(\vr,\vr_1;\omega) P(\vr_2,\vr';\omega)  P(\vr',\vr;\tau_{\rm E}) e^{i\omega\tau_{\rm E}/\hbar}\nonumber\\ && \left. \mbox{}
  + P(\vr,\vr_1;\omega)\mathcal{D}'_{\omega} P(\vr_2,\vr';\omega)  P(\vr',\vr;\tau_{\rm E}) e^{i\omega\tau_{\rm E}/\hbar}
  \vphantom{\int}
  \right\},
\end{eqnarray}
\end{widetext}
with the short-hand notations $\mathcal{D}_{\omega}=(D\Delta_{\vr}+i\omega/\hbar)$ and $\mathcal{D}'_{\omega}=(D\Delta_{\vr'}+i\omega/\hbar)$. The technical advantage of the structure of Eq.\ \eqref{eq:Kdiff} is that each term contains the same diffusive propagators. Performing several partial integrations, using $\Delta P_{\rm in} (\vr)=\Delta P_{\rm out}(\vr)=0$, $\vnabla P_{\rm in}(\vr)=-\vnabla P_{\rm out}(\vr)=-\tfrac{1}{L} \vec{e_x}$, as well as ${\cal D}_{\omega}\int_{\tau_{\rm E}}^{\infty}dt P(\vr',\vr;t) e^{i\omega t/\hbar}=-P(\vr',\vr;\tau_{\rm E}) e^{i\omega\tau_{\rm E}/\hbar}$, we are able to simplify the expression and finally obtain
\begin{align}
 \label{eq:Kdiff1}
  K (\vr_1,\vr_2;\omega)=&-\frac{4D^2}{L^2} \int d\vr d\vr' P(\vr,\vr_1;\omega) P(\vr_2,\vr';\omega) \nonumber\\
        &\times\partial_x \partial_{x'} \int_{\tau_{\rm E}}^{\infty} dt P(\vr',\vr;t) e^{i\omega t/\hbar}.
\end{align}
Together, Eq.\ \eqref{eq:dGAA-1} and \eqref{eq:Kdiff1} represent the main result for the interaction correction in antidot arrays. For zero Ehrenfest time, the time integral of Eq.\ \eqref{eq:Kdiff1} equals $P(\vr',\vr;\omega)$ and one recovers the results for quantum impurities, obtained by standard diagrammatic perturbation theory  (see Ref.\ \onlinecite{AltshulerAronovZyuzin}, where the symbol $F$ of this reference equals $F/2=-F_0^{\sigma}/(1+F_0^{\sigma})$ and the reference misses a factor two for the triplet contribution). If the Ehrenfest time is finite, it poses a short-time threshold and only electrons with a travel time larger than $\tau_{\rm E}$ contribute to the interaction correction.
We now discuss Ehrenfest-time dependence of $\delta G_{\rm AA}$ in detail for a quasi-1d and a 2d antidot array.

\subsection{Quasi-one dimensional antidot array}
\label{sec:1d}

For a quasi-1d antidot array (width $W$ much smaller than length $L$), we may simplify the diffusion propagator by taking only the diffusion mode with zero transverse momentum into account. After insertion of the diffusion propagators and the interaction into Eqs.\ \eqref{eq:dGAA-1} and \eqref{eq:Kdiff1}, and using the residue technique for the $\omega$-integration, we find
\begin{align}
 \label{eq:deltaGsum}
  \delta G_{\rm AA}=&-\frac{e^2}{h}  \sum_{m=1}^{\infty} \sum_{n=1}^{\infty}  e^{-n\frac{\tau_{\rm E}}{\tau_T}}  e^{-m^2 \frac{\tau_{\rm E}}{\tau_{\rm D}}}\nonumber \\
  &\times\left[n \frac{\tau_{\rm E} \tau_{\rm D}}{\tau_T^2} g_{m}(n \tau_{\rm D}/\tau_T) -n\frac{\tau_{\rm D}^2}{\tau_T^2} g'_{m}(n \tau_{\rm D}/\tau_T)   \right],
 \end{align} 
 where the time $\tau_T=\hbar/2\pi T$ is the inverse temperature, $\tau_{\rm D}=L^2/D\pi^2$ is the diffusion time, and the function $g_m(x)$ is expressed as
  \begin{align}
   \label{eq:gm}
   g_{m}(x)=&\frac{128}{\pi^4}\sum_{k=1}^{\infty} c_{km}^2 \frac{1}{(k^2 +x) (m^2+x)} \nonumber\\
    &\times\left\lbrace\frac{1}{k^2}+\frac{3F_0^{\sigma}}{k^2(1+F_0^{\sigma})+x}\right\rbrace,
  \end{align}
with
  \begin{equation}
  c_{km}=\begin{cases}
          {km}/({k^2-m^2}) & \mbox{if $k+m$ odd},\\
           0 & \text{else}.
         \end{cases}
 \end{equation}
 The summation over $k$ in Eq.\ \eqref{eq:gm} can be written in a closed form (see Eq.\ \eqref{eq:gmexpl} in the appendix).
We will now discuss the dependence of $\delta G_{\rm AA}$ on temperature, system size and Ehrenfest time several limiting cases.

\subsubsection{${\bm \tau}_{\rm D}\gg{\bm \tau}_T,{\bm \tau}_{\rm E}$}

\begin{figure}[t]
\includegraphics[width=2.9in]{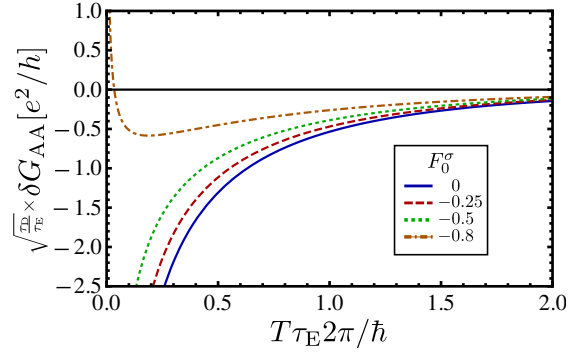}
\caption{(Color online) Interaction correction to the conductance of a quasi-one dimensional antidot array in the large-system-size regime $\tau_{\rm D}\gg\tau_T$, $\tau_{\rm E}$.}
\label{fig:Plotd1}
\end{figure}

We first consider the limit $\tau_{\rm D} \to \infty$ corresponding to a large antidot array. In this case, the Ehrenfest-time dependence of the interaction correction $\delta G_{\rm AA}$ is determined by the ratio $\tau_{\rm E}/\tau_{T}$. For $\tau_{\rm E}/\tau_{T} \ll 1$ one finds the result
\begin{equation}
 \delta G_{\rm AA}=-\frac{e^2}{h} \sqrt{\frac{\tau_T}{\tau_{\rm D}}} \frac{3\zeta(\tfrac{3}{2})}{\pi} \left[1+3\frac{2+F_0^{\sigma}-2\sqrt{1+F_0^{\sigma}}}{F_0^{\sigma}}\right],
\end{equation}
independent of $\tau_{\rm E}$ and known from diagrammatic perturbation theory. Here $\zeta(3/2) \approx 2.61238$ is the Riemann zeta function (see Appendix\ \ref{sec:appdiscussion} for details). On the other hand, for large Ehrenfest times or, equivalently, higher temperatures, $\tau_{\rm E}/\tau_{T} \gg 1$, the interaction correction $\delta G_{\rm AA}$ acquires an exponential dependence on temperature $\propto e^{-2\pi T\tau_{\rm E}/\hbar}$, 
\begin{equation}
 \delta G_{\rm AA}=-\frac{e^2}{h} \frac{4}{\pi^{3/2}} \sqrt{\frac{\tau_{\rm E}}{\tau_{\rm D}}} e^{-\frac{\tau_{\rm E}}{\tau_T}},
\end{equation}
independent of $F_0^{\sigma}$. The crossover between these two limiting cases is shown in Fig.\ \ref{fig:Plotd1} for different values of the Fermi-Liquid interaction constant $F_0^{\sigma}$.

We emphasize the influence of the interaction constant in the triplet channel $F_0^{\sigma}$: 
While for small values of $F_0^{\sigma}$, $\delta G_{\rm AA}$ is always negative and monotonously decaying as temperature is increased, a more interesting behaviour is observed at large interaction strength: At small Ehrenfest time and $F_0^{\sigma}<-\frac{3}{4}$ the contribution from the triplet channel dominates and gives rise to a positive sign of the interaction correction. On the contrary, if the Ehrenfest time is large, the prefactor of the exponential behaviour shows no dependence on $F_0^{\sigma}$ to leading order in $\frac{\tau_T}{\tau_{\rm E}}$ and is therefore always negative. Hence, at sufficiently large interaction strengths, one observes a sign change of the interaction correction, as the temperature is varied. 

\subsubsection{${\bm\tau}_T\gg{\bm \tau}_{\rm D},{\bm \tau}_{\rm E}$}

\begin{figure}[t]
\includegraphics[width=2.9in]{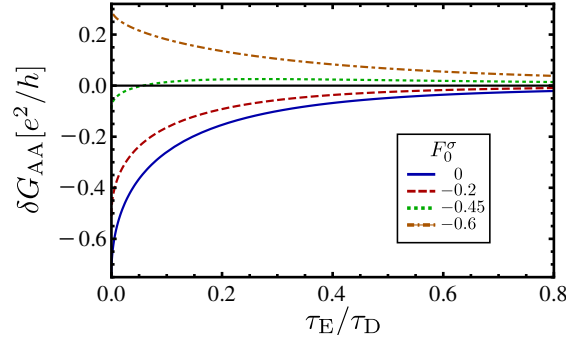}
\caption{(Color online) Interaction correction to the conductance of a quasi-one dimensional antidot array in the low-temperature regime $\tau_T\gg\tau_{\rm D}$, $\tau_{\rm E}$.}
\label{fig:PlotL}
\end{figure}

In the limit of zero temperature $\delta G_{\rm AA}$ is a function of the ratio $\tau_{\rm E}/\tau_{\rm D}$ only (we again refer to Appendix \ref{sec:appdiscussion} for details). At zero Ehrenfest time, we have 
\begin{equation}
  \delta G_{\rm AA}=- \frac{e^2}{h}
  \sum_{m=1}^{\infty} \int_0^{\infty} dx g_m(x),
\end{equation}
with $g_m(x)$ defined in Eq.\ (\ref{eq:gm}). For small $F_0^{\sigma}$, we have $\delta G_{\rm AA} \approx -(e^2/h)(0.74+F_0^{\sigma})$. In this parameter range
the singlet contribution is dominant, which leads to a negative sign of the interaction correction, and to a monotonous decay as a function of $\frac{\tau_{\rm E}}{\tau_{\rm D}}$. At larger $F_0^{\sigma}$ the triplet contribution competes with the singlet contribution, resulting in a sign change of the interaction correction for sufficiently strong interactions, starting at $F_0^{\sigma}\approx-0.5$. For $\tau_{\rm E}\gg\tau_{\rm D}$, we find an exponential dependence of $\delta G_{\rm AA}$ on $\frac{\tau_{\rm E}}{\tau_{\rm D}}$
\begin{equation}
   \delta G_{\rm AA}=-\frac{e^2}{h} \left(1+\frac{3 F_0^{\sigma}}{1+F_0^{\sigma}}\right)\frac{16(5\pi^2-48)}{3\pi^4}\frac{\tau_{\rm D}}{\tau_{\rm E}}e^{-\frac{\tau_{\rm E}}{\tau_{\rm D}}}.
\end{equation}
The crossover between the limits $\tau_{\rm E} \ll \tau_{\rm D}$ and $\tau_{\rm E} \gg \tau_{\rm D}$ is shown in Fig.\ \ref{fig:PlotL} for several representative values of $F_0^{\sigma}$. 

\subsubsection{${\bm \tau}_{\rm E}\gg{\bm \tau}_{\rm D},{\bm \tau}_T$}
Finally, in the ``classical'' limit of large Ehrenfest times, we have
\begin{equation}
  \delta G_{\rm AA}=-\frac{e^2}{h}  \tfrac{\tau_{\rm E} \tau_{\rm D}}{\tau_T^2} e^{-\frac{\tau_{\rm E}}{\tau_T}}  e^{- \frac{\tau_{\rm E}}{\tau_{\rm D}}} g_{1}(\tfrac{\tau_{\rm D}}{\tau_T}),
 \end{equation} 
where $g_1$ is defined in Eq.\ (\ref{eq:gm}). In this parameter regime the interaction correction has the characteristic exponential suppression $\delta G_{\rm AA} \propto e^{-2\pi T\tau_{\rm E}/\hbar - {\tau_{\rm E}}/{\tau_{\rm D}}}$.

\subsection{Large 2d system}
\label{sec:2d}
\begin{figure}[t]
\includegraphics[width=2.9in]{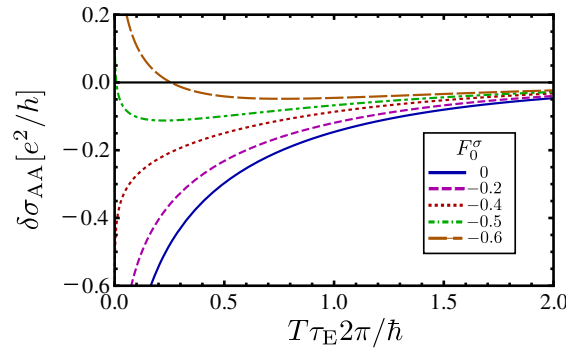}
\caption{(Color online) Interaction correction to the conductivity $\sigma$ of a two-dimensional antidot array for various values of the interaction strength $F_0^{\sigma}$ in the triplet channel.}
\label{fig:Plotd2}
\end{figure}

We will now consider a two-dimensional antidot array of dimensions $L \times W$, where we restrict ourselves to the limit of large system size, $\tau_{\rm D} \gg \tau_{\rm E}$, $\tau_{T}$. In the large-size limit, the relevant quantity is the conductivity $\sigma = G L/W$ (although we here write results for the conductivity, we formally calculate the conductance and multiply with the geometrical factor $L/W$, see also the discussion in Sec.\ \ref{sec:skeleton}). We may then express Eqs.\ \eqref{eq:dGAA-1} and \eqref{eq:Kdiff1} in momentum space,
\begin{align}
 \label{eq:dG2d}
  \delta \sigma_{\rm AA}=&-\frac{4 \nu e^2 D}{\pi\hbar^2}  \int d\omega\partial_{\omega}\left(\omega \coth\frac{\omega}{2 T}\right)\nonumber\\
   &\times \mathrm{Im}\left\lbrace \int \frac{d^2\vq}{(2\pi)^2} U^{\rm R}(\vq;\omega) \frac{D q_x^2 e^{i\omega\tau_{\rm E}}e^{-D\vq^2\tau_{\rm E}}}{(D\vq^2-i\omega/\hbar)^3}\right\rbrace.
\end{align}
In the limit of zero Ehrenfest time, this expression simplifies to the well-known result of diagrammatic perturbation theory.\cite{AleinerAG,ZalaNarozhnyAleiner} The full Ehrenfest-time dependence is shown in Fig.\ \ref{fig:Plotd2}. For $\tau_{\rm E}\ll\tau_T$, we have the asymptotic behavior
\begin{equation}
\delta \sigma_{\rm AA} =-\frac{e^2}{\pi h} \left[1+3\left(1-\frac{\ln(1+F_0^{\sigma})}{F_0^{\sigma}}\right)\right]\ln\frac{\tau_T}{\tau_{\rm E}},
\end{equation}
which coincides with the well-known expression of quantum impurities, where the role of the elastic scattering time as a short-time cutoff is taken over by the Ehrenfest time. In the opposite limit $\tau_{\rm E}\gg\tau_T$, we obtain an exponential dependence on temperature,
\begin{equation}
\delta \sigma_{\rm AA} =-\frac{e^2}{\pi h} e^{-\frac{\tau_{\rm E}}{\tau_T}}.
\end{equation}
As in the one-dimensional situation, at small $F_0^{\sigma}$ the singlet contribution dominates the interaction correction, while at larger $F_0^{\sigma}$ the triplet contribution competes, and a sign change of the interaction correction as a function of $\frac{\tau_{\rm E}}{\tau_T}$ is observed if $F_0^{\sigma}\lesssim-0.45$. 

\section{Conclusion}
\label{sec:conclusion}
In this article, we considered the effect of a finite Ehrenfest time on the interaction correction of a conductor in which the motion of the electrons is described by chaotic classical dynamics. Using semiclassical theory of transport, we derived an expression for the interaction correction containing only the interaction propagator and coarse-grained classical propagators of the electronic motion. We confirm the result of Ref.\ \onlinecite{BrouwerKupferschmidt}, obtained for a double ballistic quantum dot, that the Ehrenfest time enters as a short-time threshold for the interaction correction. In other words, the minimal time it takes to traverse system for trajectories responsible for the interaction corrections is the Ehrenfest time.

As a specific and experimentally relevant example we applied the formalism to antidot arrays, where the coarse-grained classical dynamics follows a diffusion equation. At zero Ehrenfest time, we recovered the well known results of the diagrammatic perturbation theory for a disordered metal.\cite{AltshulerAronov} If the Ehrenfest time is large, we found that the interaction correction is exponentially suppressed $\propto e^{-{\tau_{\rm E}}/{\tau_{\rm D}}} e^{-2\pi T\tau_{\rm E}/\hbar}$. While the factor $e^{-{\tau_{\rm E}}/{\tau_{\rm D}}}$ is also present for weak localization, the suppression with temperature is specific to the interaction correction. Unlike the dwell time $\tau_{\rm D}$, which governs the Ehrenfest-time dependence of weak localization, temperature is a variable that can be easily controlled experimentally without changing the classical dynamics, making the interaction correction a promising experimental signature of the Ehrenfest-time dependence of quantum transport. (We note that weak 
localization depends on temperature implicitly via its dependence on the dephasing time. However, an independent measurement of the dephasing time that enters into the expression for the Ehrenfest-time dependence of the weak localization correction is problematic.\cite{kn:tian2007})

A particular signature of the underlying classical motion is a sign change of the interaction corrections for strong enough interactions. Associated with this sign change is a non-monotonous temperature dependence of the interaction correction, in the temperature range $T \sim \hbar/\tau_{\rm E}$. As long as only the Fock contribution is considered, the sign of the interaction correction is negative. If the Hartree contribution --- more precisely, the triplet channel of interaction --- is added, there is a competition between Hartree and Fock-type corrections, and the sign of the total interaction correction at small Ehrenfest time may change as a function of the interaction strength. For large Ehrenfest time, the Fock contribution always dominates, so that the sign of the interaction correction is independent on the interaction strength in that limit.

As mentioned above, the sign change of the interaction corrections for systems with small Ehrenfest times requires a rather strong interaction in the triplet channel. In particular, in $2d$ systems the threshold was estimated to be $F_0^\sigma\lesssim -0.45$, where $F_0^\sigma$ is the corresponding Fermi liquid constant, whereas in quasi one-dimensional systems the condition reads $F_0^\sigma\lesssim -0.75$. Let us focus on $2d$ systems with Coulomb interaction, for which the condition is less restrictive. Both numerical and experimental results for $F_0^\sigma$ are available in the literature. In general, $F_0^\sigma$ is a function of the gas parameter $r_s=\sqrt{2}e^2/\varepsilon \hbar v_F$, where $\varepsilon$ is the static dielectric constant and $v_F$ the Fermi velocity. To get an estimate for typical values of $r_s$ to be expected in antidot array experiments, we take $v_F\approx 3\times 10^5 m/s$, which was reported in Ref.\ \onlinecite{Yevtushenko} for antidot arrays fabricated from GaAs/
AlGaAs heterostructures. Together with the dielectric constant for GaAs $\varepsilon\approx 13$ we obtain $r_s\approx 0.8$. In Ref.~\onlinecite{Minkov2006} the constant $F_0^\sigma$ for a given gas parameter $r_s$ was extracted from experimental data using the results for the interaction corrections to conductivity\cite{Altshuler85, Finkelstein90, ZalaNarozhnyAleiner,GornyiMirlin}. For systems with $r_s\approx 1$, typical values of $F_0^\sigma$ were found to be of the order of $-0.35$. This seems consistent with numerical results obtained in Ref.~\onlinecite{Kwon1994}, were systems with moderately large $r_s$ were analyzed. For the maximal value of $r_s=5$ considered in this paper, the Fermi liquid constant decreased further down to $F_0^\sigma=-0.5$. Considerably larger negative values up to $F_0^\sigma \approx -0.7$ were inferred for systems with $r_s\approx 22$ in Ref.~\onlinecite{Noh2003}. We take this as evidence that $2d$ systems with sufficiently strong triplet channel interactions are 
realizable, provided the additional antidot structure may be superimposed. Relevant values of the gas parameter $r_s$ are likely in the range of $r_s\approx 3-5$.

\acknowledgments
We gratefully acknowledge discussions with Maxim Breitkreiz, Alexander Finkel'stein, Igor Gornyi and Alexander Mirlin. This work is supported by the Alexander von Humboldt Foundation in the framework of the Alexander von Humboldt Professorship, endowed by the Federal Ministry of Education and Research and by the German Research Foundation (DFG) in the framework of the Priority Program 1459 ``Graphene''.

\appendix

\begin{widetext}

\section{Details of the semiclassical calculation}

\subsection{Sum rule}
\label{sec:sumrule}

Here we show how to derive the sum rule Eq.\ \eqref{eq:sumrule}.
We start by noticing that
\begin{equation}
 A_{\alpha}^2=\left|\det\matr{\frac{\partial^2 \mathcal{S}_{\alpha}}{\partial\vr'\partial\vr}}{\frac{\partial^2 \mathcal{S}_{\alpha}}{\partial \vr' \partial\xi}}{\frac{\partial^2 \mathcal{S}_{\alpha}}{\partial\xi\partial \vr}}{\frac{\partial^2 \mathcal{S}_{\alpha}}{\partial \xi\partial \xi}}\right|=\left|\det\matr{\frac{\partial \vp'_{\alpha}}{\partial\vr}}{\frac{\partial \vp'_{\alpha}}{\partial\xi}}{\frac{\partial \tau_{\alpha}}{\partial \vr}}{\frac{\partial \tau_{\alpha}}{\partial \xi}}\right|.
\end{equation}
Hence we may write
\begin{eqnarray}
  \sum_{\alpha:\vr'\rightarrow \vr;\xi} A_{\alpha}^2 f(\vp'_{\alpha},\vp_{\alpha},\tau_{\alpha})
  &=& \int_{0}^{\infty} dt \int d\vp' \sum_{\alpha:\vr'\rightarrow \vr;\xi} \delta(t-\tau_{\alpha})\delta(\vp'-\vp'_{\alpha}) f(\vp',\vp_{\alpha},t) \left|\det\matr{\frac{\partial\vr}{\partial\vp'}}{\frac{\partial \xi}{\partial\vp'}}{\frac{\partial\vr}{\partial t}}{\frac{\partial \xi}{\partial t}} \right|^{-1}.
\end{eqnarray}
The determinant serves as a Jacobian for the transformation $(t,\vp')\rightarrow(\xi,\vr)$,
\begin{eqnarray}
  \sum_{\alpha:\vr'\rightarrow \vr;\xi} \delta(t-\tau_{\alpha})\delta(\vp'-\vp'_{\alpha}) f(\vp',\vp_{\alpha},t) \left|\det\matr{\frac{\partial\vr}{\partial\vp'}}{\frac{\partial \xi}{\partial\vp'}}{\frac{\partial\vr}{\partial t}}{\frac{\partial \xi}{\partial t}} \right|^{-1} &=&
  \delta[\xi-H(\vr,\vp(\vr',\vp';t))]\delta[\vr-\vr(\vr',\vp';t)]
  \nonumber \\ && \mbox{} \times
  f[\vp',\vp(\vr',\vp';t),t],
\end{eqnarray}
where $(\vr(\vr',\vp';t),\vp(\vr',\vp';t))$ is the phase space point that a trajectory originating from $(\vr',\vp')$ reaches after time $t$.
After insertion of $\int d\vp \delta(\vp-\vp(\vr',\vp';t))$ we finally arrive at Eq.\ \eqref{eq:sumrule}.

\subsection{Convolution rule}
\label{sec:convolutionrule}
In this appendix we derive the convolution rule
\begin{eqnarray}
 \label{eq:convolution-app}
  \lefteqn{\int d\vr_1 d\vr_2  {\cal G}^{\rm A}(\vr',\vr_2;\xi) {\cal G}^{\rm A}(\vr_2,\vr_1;\xi-\omega) {\cal G}^{\rm A}(\vr_1,\vr;\xi)  f(\vr_1-\vr_2)} \nonumber\\
 &=&-\frac{1}{\hbar^2} \frac{2\pi}{(-2\pi i\hbar)^{3/2}} \sum_{\alpha:\vr'\rightarrow\vr;\xi} A_{\alpha} e^{- i \mathcal{S}_{\alpha}/\hbar } \int_{0}^{\tau_{\alpha}} dt \int_0^{t} dt' f(\vr_{\alpha}(t)-\vr_{\alpha}(t')) e^{i\omega(t-t')/\hbar},
\end{eqnarray}
where $f(\vr)$ is an arbitrary function. (Here we omitted contributions from stationary configurations of trajectories, that cannot be connected to a single trajectory. For the calculation of the interaction correction, such contributions drop out upon pairing with the retarded trajectory.)

To this end, we first prove an auxiliary identity for the stability amplitudes. Hereto we consider a trajectory $\alpha$ that connects $\vr'$ with $\vr$. Further, let $\vr_1$ be a point on trajectory $\alpha$, and $\alpha'$ ($\alpha''$) be the part of trajectory $\alpha$ connecting $\vr'$ with $\vr_1$ ($\vr_1$ with $\vr$). The stability amplitude $A_{\alpha}$ can be then written as\cite{Gutzwiller}
\begin{equation}
 A_{\alpha}^2=\frac{1}{v_{\rm F}^2}\left|\frac{\partial^2 \mathcal{S}_{\alpha}(\vr,\vr')}{\partial r_{\perp} \partial r'_{\perp}}\right|=\frac{1}{v_{\rm F}^2}\left|\frac{\partial p'_{\perp\alpha}}{\partial{r_{\perp}}}\right|, \quad  p'_{\perp\alpha} (\vr,\vr')=-\frac{\partial \mathcal{S}_{\alpha}(\vr,\vr')}{\partial{r'_{\perp}}},
\end{equation}
 where $v_{\rm F}$ is the Fermi velocity and $r_{\perp}$ ($r'_{\perp}$) denote displacements perpendicular to the trajectory $\alpha$. The last equation implicitly defines $r_{\perp\alpha}(r'_{\perp},p'_{\perp})$. We then introduce 
\begin{equation}
 B_{\alpha}=-\left(\frac{\partial r_{\perp\alpha}}{\partial{p'_{\perp\alpha}}}\right)_{r'_{\perp}}, \quad  B_{\alpha'}=-\left(\frac{\partial r_{\perp 1\alpha'}}{\partial{p'_{\perp\alpha'}}}\right)_{r'_{\perp}}, \quad  B_{\alpha''}=-\left(\frac{\partial r_{\perp\alpha''}}{\partial{p_{\perp 1\alpha''}}}\right)_{r_{\perp 1}},
\end{equation}
such that $A_{\alpha}^2=v_{\rm F}^{-2}\left|B_{\alpha}^{-1}\right|$, $A_{\alpha'}^2=v_{\rm F}^{-2}\left|B_{\alpha'}^{-1}\right|$, $A_{\alpha''}^2=v_{\rm F}^{-2}\left|B_{\alpha''}^{-1}\right|$. Then the following identity holds:
 \begin{equation}
  \label{eq:convA}
  B_{\alpha}= B_{\alpha'} B_{\alpha''}\left(\frac{\partial^2\mathcal{S}_{\alpha'}(\vr_1,\vr')}{\partial r_{\perp 1}^2}+\frac{\partial^2\mathcal{S}_{\alpha''}(\vr,\vr_1)}{\partial r_{\perp 1}^2}\right).
 \end{equation}
 
For the proof of Eq.\ (\ref{eq:convA}), we note that $B_{\alpha}$ measures the change of the final coordinate of $\alpha$ induced by a small change of the initial momentum. When we consider $\alpha$ to be composed by $\alpha'$ and $\alpha''$, a small change of the initial momentum leads to a change of the intermediate coordinate and momentum, which results in a change of the final coordinate,
\begin{align}
 &\left(\frac{\partial r_{\perp \alpha}}{\partial p'_{\perp \alpha}}\right)_{r'_{\perp}}\nonumber\\
=&\left(\frac{\partial r_{\perp \alpha''}}{\partial p_{\perp 1\alpha''}}\right)_{r_{\perp 1}}
\left(\frac{\partial p_{\perp 1\alpha'}}{\partial p'_{\perp \alpha'}}\right)_{r'_{\perp}}
+\left(\frac{\partial r_{\perp \alpha''}}{\partial r_{\perp 1\alpha''}}\right)_{p_{\perp 1}}
\left(\frac{\partial r_{\perp 1\alpha'}}{\partial p'_{\perp \alpha'}}\right)_{r'_{\perp}}\nonumber\\
 =&\left(\frac{\partial r_{\perp \alpha''}}{\partial p_{\perp 1\alpha''}}\right)_{r_{\perp 1}}
   \left(\frac{\partial r_{\perp 1\alpha'}}{\partial p'_{\perp \alpha'}}\right)_{r'_{\perp}}
\left[\left(\frac{\partial p'_{\perp \alpha'}}{\partial r_{\perp 1\alpha'}}\right)_{r'_{\perp}}
\left(\frac{\partial p_{\perp 1\alpha'}}{\partial p'_{\perp \alpha'}}\right)_{r'_{\perp}}
+\left(\frac{\partial r_{\perp \alpha''}}{\partial r_{\perp 1\alpha''}}\right)_{p_{\perp 1}}
\left(\frac{\partial p_{\perp 1\alpha''}}{\partial r_{\perp \alpha''}}\right)_{r_{\perp 1}}\right]\nonumber\\
 =&\left(\frac{\partial r_{\perp \alpha''}}{\partial p_{\perp 1\alpha''}}\right)_{r_{\perp 1}}
   \left(\frac{\partial r_{\perp 1\alpha'}}{\partial p'_{\perp \alpha'}}\right)_{r'_{\perp}}
    \left[\left(\frac{\partial p_{\perp 1\alpha'}}{\partial r_{\perp 1\alpha'}}\right)_{r'_{\perp}}
    -\left(\frac{\partial p_{\perp 1\alpha''}}{\partial r_{\perp 1\alpha''}}\right)_{r_{\perp}}\right]
\end{align}
The last line yields Eq.\ \eqref{eq:convA}.

In a similar fashion, one verifies that
\begin{equation}
  \label{eq:convA2}
  B_{\alpha'}= B_{\alpha} B_{\alpha''}\left(\frac{\partial^2\mathcal{S}_{\alpha''}(\vr,\vr_1)}{\partial r_{\perp}^2}-\frac{\partial^2\mathcal{S}_{\alpha}(\vr,\vr')}{\partial r_{\perp}^2}\right).
 \end{equation}

The identity we need for the derivation of the convolution rule (\ref{eq:convolution-app}) involves the partitioning of a single trajectory $\alpha$ into three trajectories $\alpha'$ ($\vr'\rightarrow\vr_1$), $\alpha''$ ($\vr_1\rightarrow\vr_2$) and $\alpha'''$ ($\vr_2\rightarrow\vr$). In this case, we have
\begin{equation}
  \label{eq:convA3}
  B_{\alpha}= B_{\alpha'} B_{\alpha''} B_{\alpha'''}\left[\left(\frac{\partial^2\mathcal{S}_{\alpha'}}{\partial r_{\perp 1}^2}+\frac{\partial^2\mathcal{S}_{\alpha''}}{\partial r_{\perp 1}^2}\right)\left(\frac{\partial^2\mathcal{S}_{\alpha''}}{\partial r_{\perp 2}^2}+\frac{\partial^2\mathcal{S}_{\alpha'''}}{\partial r_{\perp 2}^2}\right)-\left(\frac{\partial^2\mathcal{S}_{\alpha''}}{\partial r_{\perp 1}\partial r_{\perp 2}}\right)^2\right].
 \end{equation}
To see this, one introduces the trajectory $\tilde{\alpha}$ as connection of $\alpha$ and $\alpha'$ and makes use of Eqs.\ \eqref{eq:convA} and \eqref{eq:convA2}.

We now turn to the proof of the convolution rule (\ref{eq:convolution-app}). Hereto, we define
\begin{align}
 K(\vr,\vr';\omega)=&\int d\vr_1 d\vr_2 \delta(\vr_1-\vr_2-\va) {\cal G}^{\rm A}(\vr',\vr_2;\xi) {\cal G}^{\rm A}(\vr_2,\vr_1;\xi-\omega) {\cal G}^{\rm A}(\vr_1,\vr;\xi),
\end{align}
where $\va$ is arbitrary, but fixed. 
With the abbreviation $\tilde{\vr}_1=\vr_1-\va$ we write
\begin{equation}
 K(\vr,\vr';\omega)=\int d\vr_1 {\cal G}^{\rm A}(\vr',\tilde{\vr}_1;\xi) {\cal G}^{\rm A}(\tilde{\vr}_1,\vr_1;\xi-\omega) {\cal G}^{\rm A}(\vr_1,\vr;\xi).
\end{equation}
We then insert the semiclassical expressions for the Green functions which expresses the former equation as a sum over trajectories $\alpha'$ (from $\vr'$ to $\tilde{\vr}_1$),  $\alpha''$ (from $\tilde{\vr}_1$ to $\vr_1$), and $\alpha'''$ (from $\vr_1$ to $\vr$).

The integration over $\vr_1$ is carried out within stationary phase approximation. Here we only take into account stationary phase configurations, where $\alpha'$, $\alpha''$ and $\alpha'''$ are connected to a single trajectory. Other configurations play no role for the calculation of the conductance, since the three advanced trajectories are paired with a single retarded trajectory. Hence the convolution $K$ may be written as a sum over trajectories $\alpha$ connecting $\vr'$ with $\vr$ and a sum over points $\vr_1^{(0)}$, for which $\alpha$ first passes through $\tilde{\vr}_1^{(0)}=\vr_1^{(0)}-\va$ and then through $\vr_1^{(0)}$. Deviations $\Delta \vr_1=(\Delta x_1,\Delta y_1)$ from $\vr_1^{(0)}$ may be parametrized as
\begin{align}
 \label{eq:trafo}
 \Delta r_{\perp 1}&=- \sin(\theta)\Delta x_1+ \cos(\theta)\Delta y_1\nonumber\\
 \Delta \tilde{r}_{\perp 1}&=- \sin(\tilde{\theta})\Delta x_1+ \cos(\tilde{\theta})\Delta y_1
\end{align}
where $\theta$ ($\tilde{\theta}$) is the angle of the momentum of trajectory $\alpha$ at $\vr_1^{(0)}$ ($\tilde{\vr}_1^{(0)}$), and in turns $\Delta r_{\perp 1}$ ($\Delta \tilde{r}_{\perp 1}$) represent perpendicular displacements of trajectory $\alpha$ at $\vr_1^{(0)}$ ($\tilde{\vr}_1^{(0)}$). We then expand the sum of the actions of trajectories $\alpha'$, $\alpha''$ and $\alpha'''$ up to second order in $\Delta \vr_1$:
\begin{align}
 &\mathcal{S}_{\alpha'}(\tilde{\vr}_1,\vr';\xi)+ \mathcal{S}_{\alpha''}(\vr_1,\tilde{\vr}_1;\xi-\omega)+ \mathcal{S}_{\alpha'''}(\vr,\vr_1;\xi)=\mathcal{S}_{\alpha}(\vr,\vr';\xi)-\omega\tau+\Delta \mathcal{S}(\Delta \vr_1),
\end{align}
where $\tau$ is the duration of $\alpha$ between $\tilde{\vr}_1^{(0)}$ and $\vr_1^{(0)}$, and $\Delta \mathcal{S}(\Delta \vr_1)$ is given by
\begin{align}
 \Delta \mathcal{S}(\Delta \vr_1)=&\frac{1}{2}\left[\frac{\partial^2\mathcal{S}_{\alpha'}(\tilde{\vr}_1^{(0)},\vr')}{\partial \tilde{r}_{\perp 1}^2}+\frac{\partial^2\mathcal{S}_{\alpha''}(\vr_1^{(0)},\tilde{\vr}_1^{(0)})}{\partial \tilde{r}_{\perp 1}^2}\right] \Delta \tilde{r}_{\perp 1}^2+\frac{1}{2}\left[\frac{\partial^2\mathcal{S}_{\alpha''}(\vr_1^{(0)},\tilde{\vr}_1^{(0)})}{\partial r_{\perp 1}^2}+\frac{\partial^2\mathcal{S}_{\alpha'''}(\vr,\vr_1^{(0)})}{\partial r_{\perp 1}^2}\right] \Delta r_{\perp 1}^2\nonumber\\
 &+\left[\frac{\partial^2\mathcal{S}_{\alpha''}(\vr_1^{(0)},\tilde{\vr}_1^{(0)})}{\partial \tilde{r}_{\perp 1} \partial r_{\perp 1}}\right]\Delta r_{\perp 1}  \Delta \tilde{r}_{\perp 1},
\end{align}
where as a consequence of energy conservation only perpendicular displacements need to be considered. The integration over $\Delta \vr_1$ can then be accomplished and using Eq. \eqref{eq:convA3} we get
\begin{equation*}
 K(\vr,\vr';\omega)=\left(\frac{2\pi}{(2\pi i\hbar)^{3/2}}\right)^3 (2\pi i\hbar)\sum_{\alpha:\vr'\rightarrow \vr;\xi}\sum_{\lbrace \vr_1^{(0)}\rbrace}\frac{1}{v_{\rm F}^2}\frac{1}{|\sin(\theta-\tilde{\theta})|}A_{\alpha} 
 e^{-i \left(\mathcal{S}_{\alpha} - \omega \tau \right)/\hbar}
\end{equation*}
where the factor $|\sin(\theta-\tilde{\theta})|^{-1}$ originates from the Jacobian of the transformation \eqref{eq:trafo}. (A possible phase shift from taking the squareroot of Eq.\ \eqref{eq:convA3} is needed to restore the correct Maslov index. For our calculation however, the Maslov index plays no role and we drop it in our expressions.)

On the other hand, we have
\begin{equation*}
 \int_{0}^{\tau_{\alpha}} dt \int_0^{t} dt' \delta^{(2)}(\vr_{\alpha}(t)-\vr_{\alpha}(t')-\va)=\sum_{\lbrace \vr_1^{(0)}\rbrace}\frac{1}{v_{\rm F}^2}\frac{1}{|\sin(\theta-\tilde{\theta})|}.
\end{equation*}

 With that, we finally obtain
\begin{equation}
 K(\vr,\vr',\omega)=-\frac{1}{\hbar^2} \frac{2\pi}{(2\pi i\hbar)^{3/2}} \sum_{\alpha:\vr'\rightarrow \vr;\xi} A_{\alpha} e^{-i\mathcal{S}_{\alpha}/\hbar} \int_{0}^{\tau_{\alpha}} dt \int_0^{t} dt' \delta^{(2)}(\vr_{\alpha}(t)-\vr_{\alpha}(t')-\va)  e^{i\omega(t-t')}.
\end{equation}
  Multiplying with $f(\va)$ and integrating over $\va$ then yields Eq.\ \eqref{eq:convolution-app}.

\end{widetext}

\subsection{Summation over classical trajectories involving a small-angle encounter}
\label{sec:encrules}

The summation over classical trajectories with a small-angle encounter as given in Fig.\ \ref{fig:diag2} is performed using the procedure of Refs.\ \onlinecite{MuellerNJP,Brouwer}. We here outline the main points of this calculation.

\begin{figure}[t]
\includegraphics[width=2.9in]{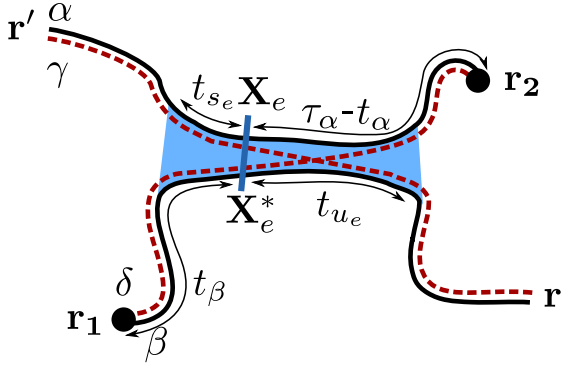}
\caption{(Color online) Schematic drawing of an encounter, formed by the trajectories $\alpha$, $\beta$, $\gamma$, and $\delta$, with timescales and phase-space points as described in the main text.}
\label{fig:encounter}
\end{figure}

The four trajectories $\alpha$ (from $\vr'$ to $\vr_2$), $\beta$ (from $\vr_1$ to $\vr$), $\gamma$ (from $\vr'$ to $\vr$), and $\delta$ (from $\vr_1$ to $\vr_2$) are piecewise paired as shown in Fig.\ \ref{fig:encounter}. We start by noting that the choice of the retarded trajectories $\alpha$ and $\beta$ fully specifies the advanced trajectories $\gamma$ and $\delta$, since the linearized chaotic dynamics allows for precisely one unique solution of a trajectory that satisfies the initial and final conditions required for the pairing shown in Fig.\ \ref{fig:encounter}. Moreover, the products of the stability amplitudes are equal, $A_{\alpha} A_{\beta} = A_{\gamma} A_{\delta}$, so that the product of four Green functions required for the calculation of $\delta G_{\rm AA}$ can be written as
\begin{eqnarray}
  \lefteqn{{\cal G}^{\rm R}(\vr,\vr_1;\xi) {\cal G}^{\rm A}(\vr_1,\vr_2,\xi-\omega) {\cal G}^{\rm R}(\vr_2,\vr';\xi)  {\cal G}^{\rm A}(\vr',\vr;\xi)} ~~~~~~~~~~~
  \nonumber\\
  &=&\frac{1}{2\pi\hbar^3} \sum_{\alpha:\vr'\to\vr_2;\xi} \sum_{\beta:\vr_1\to\vr;\xi} A_{\alpha}^2 A_{\beta}^2 
  e^{i \Delta\mathcal{S}/\hbar} e^{i\omega\tau_{\delta}/\hbar}, \nonumber \\
\end{eqnarray}
where $\Delta \mathcal{S}$ is the action difference $\mathcal{S}_{\alpha}+\mathcal{S}_{\beta}-\mathcal{S}_{\gamma}-\mathcal{S}_{\delta}$. The summation over trajectories $\alpha$ and $\beta$ is restricted to those trajectories that undergo (at least) one small-angle encounter.

The action difference $\Delta {\cal S}$ has two contributions: One contribution from the length difference of the retarded trajectories $\alpha$ and $\beta$ vs.\ the advanced trajectories $\gamma$ and $\delta$, and one contribution from the different energy $\xi - \omega$ associated with the trajectory $\delta$. The former contribution has been calculated in Refs.\ \onlinecite{TurekRichter,Spehner} and equals $s_{\rm e} u_{\rm e}$; the latter contribution equals $\omega \tau_{\delta}$, where $\tau_{\delta}$ is the duration of the trajectory $\delta$. Note that the product $s_{\rm e} u_{\rm e}$ is independent of the position of the phase space point $\vX_{\rm e}$ along the encounter. 

The trajectories $\alpha$ and $\beta$ are enumerated by first picking a phase space point $\vX_{\rm e}$ on the trajectory $\alpha$. The Poincar\'e surface of section at this point may be parameterized with stable and unstable phase space coordinates, which are denoted $s_{\rm e}$ and $u_{\rm e}$ respectively. Moving the Poincar\'e surface of section along the trajectory, the unstable (stable) coordinate grows (shrinks) as $e^{\pm \lambda t}$, where $\lambda$ is the Lyapunov coefficient. We choose the origin of the coordinate system such that the trajectory $\alpha$ pierces the Poincar\'e surface of section at coordinates $(s_{\rm e},u_{\rm e}) = (0,0)$. The point $\vX_{\rm e}$ is part of an encounter formed by trajectories $\alpha$ and $\beta$, if $\beta$ passes through the Poincar\'e surface of section at phase-space distance $|s_{\rm e}|<c$ and $|u_{\rm e}|<c$, where $c$ is a cutoff scale, below which the chaotic classical motion can be linearized. (One can always simultaneously rescale the coordinates $s$ 
and $u$, such that the cut-off scale is the same for both coordinates.) The cut-off scale $c$ enters the final results in the combination $\ln(c^2/\hbar)$ only, so that the precise value of $c$ is unimportant, as long as $c$ represents a scale characteristic of the {\em classical} dynamics. One verifies that the choice of the phase space point $\vX_{\rm e}$ on the trajectory $\alpha$ and of the phase space coordinates for the trajectory $\beta$ also specify the two remaining trajectories $\gamma$ and $\delta$. Indeed, since $\gamma$ is paired with $\alpha$ before the encounter, and with $\beta$ after the encounter, it pierces through the Poincar\'e surface of section at the same unstable coordinate as $\beta$ and the same stable coordinate as $\alpha$. Similar considerations apply to the trajectory $\delta$. The summation over trajectories $\alpha$ and $\beta$ is then written as
\begin{widetext}
\begin{eqnarray}
  \sum_{\alpha:\vr'\to\vr_2;\xi} \sum_{\beta:\vr_1\to\vr;\xi}
  A_{\alpha}^2 A_{\beta}^2 \ldots
  &=& \int d\vp_{\xi} d\vp'_{\xi} d\vp_{1,\xi} d\vp_{2,\xi}
  \int d\vX_{\rm e} \int_{-c}^{c} ds_{\rm e} du_{\rm e} 
  \int_0^{\infty} d\tau_{\alpha} \int_0^{\infty} d\tau_{\beta}
  \int_0^{\tau_{\alpha}} dt_{\alpha} \int_0^{\tau_{\beta}} dt_{\beta}
  \frac{1}{t_{\rm enc}}
  \nonumber \\ && \mbox{} \times
  \rho_{\xi}(\vr',\vp' \to \vX_{\rm e};\tau_{\alpha} - t_{\alpha})
  \rho_{\xi}(\vX_{\rm e} \to \vr_2,\vp_2;t_{\alpha})
  \nonumber \\ && \mbox{} \times
  \rho_{\xi}(\vr_1,\vp_1 \to \vX_{\rm e}^*;t_{\beta})
  \rho_{\xi}(\vX_{\rm e}^* \to \vr,\vp;\tau_{\beta} - t_{\beta}) \ldots, ~~~
\end{eqnarray}
\end{widetext}
where $\vX_{\rm e}^*$ is the phase space point located at phase-space displacement $(s_{\rm e},u_{\rm e})$ from $\vX$ and the dots indicate an arbitrary function of the end-point coordinates of the trajectories $\alpha$ and $\beta$. The time $t_{\rm enc}$ denotes the duration of the encounter. The factor $t_{\rm enc}$ in the denominator accounts for the fact, that $\vX_{\rm e}$ can be chosen anywhere along the encounter. The ends of the encounter are determined by the condition that $\max(|s|,|u|)=c$, or that one of the trajectories involved ends, whichever occurs first. Since the phase space coordinates $s$ and $u$ decreases/increase exponentially upon proceeding along the trajectories $\alpha$ and $\beta$, with a rate given by the Lyapunov exponent $\lambda$, one finds that $t_{\rm enc}$ is given by the expression
\begin{equation}
  t_{\rm enc} = \min[\lambda^{-1} \ln(c/|s_{\rm e}|),t_{\beta}] +
  \min[\lambda^{-1} \ln(c/|u_{\rm e}|),t_{\alpha}].
\end{equation}
The four different scenarios, depending on whether $t_{\beta}$ and $t_{\alpha}$ are larger or smaller than $\lambda^{-1} \ln(c/|s_{\rm e}|)$ and $\lambda^{-1} \ln(c/|u_{\rm e}|)$, respectively, correspond to the four contributions to $\delta G_{\rm AA}^{\rm F,2a}$--$\delta G_{\rm AA}^{\rm F,2d}$ to $\delta G_{\rm AA}^{\rm F,2}$. The same situation also occurs in the calculation of the shot noise, see Ref.\ \onlinecite{WhitneyJacquod,BrouwerRahavPRB06}.

The next step in the calculation is to replace the exact trajectory densities $\rho_{\xi}$ by the coarse-grained ones. In order to account for correlations inside the encounter, we introduce phase space points $\vX$ and $\vX'$ at the beginning and end of the encounter --- if the phase-space points $\vX_1$ and $\vX_2$ associated with the interaction position are not inside the encounter. Outside the encounter region we may replace the product of the exact trajectory densities by the product of the classical propagators, whereas inside the encounter only a single classical propagator remains. After coarse graining the classical propagators are insensitive to small phase-space difference between $\vX_{\rm e}$ and $\vX_{\rm e}^*$, so that we may replace $\vX_{\rm e}^*$ by $\vX_{\rm e}$. As a result, the integration over $s_{\rm e}$ and $u_{\rm e}$ and the integration over $\vX_{\rm e}$ can be performed separately. The integration over $\vX_{\rm e}$ can be performed using a convolution rule for the classical 
propagators,
\begin{equation}
 \int d\vX_{\rm e} P(\vX_2,\vX_{\rm e};t_2) P(\vX_{\rm e},\vX_1;t_1)=P(\vX_2,\vX_1;t_1+t_2).
\end{equation}
The contribution $\delta G_{\rm AA}^{\rm F,2a}$ is then expressed as Eq.\ \eqref{eq:dGAA-F1} with $K^1$ replaced by
\begin{eqnarray}
 \label{eq:K2a}
  \lefteqn{K^{2a}(\vX_1,\vX_2;\omega)} \nonumber \\
 &=&-\int d\vX d\vX' P_{\rm in}(\vX) P(\vX,\vX_1;\omega) P(\vX_2,\vX';\omega) P_{\rm out}(\vX') \nonumber\\ 
 && \mbox{} \times\frac{1}{2\pi\hbar}  \int_{-c}^c ds du P(\vX',\vX;t_{\rm enc}) e^{i\omega t_{\rm enc}/\hbar}\frac{e^{isu/\hbar}}{t_{\rm enc}},
\end{eqnarray}
where $P(\vX,\vX';\omega)$ is the Fourier transform of $P(\vX,\vX';t)$, see Eq.\ (\ref{eq:FT}). The encounter time for this contribution is given by $t_{\rm enc} = \lambda^{-1} \ln(c^2/|su|)$. In order to perform the integration over the phase-space coordinate $s$ and $u$ we make use of the integral identity
\begin{equation}
  \label{eq:enc1}
  \frac{1}{2\pi\hbar}\int_{-c}^{c} ds du \frac{e^{isu/\hbar}}{t_{\rm enc}} f(t_{\rm enc})=\frac{\partial}{\partial \tau_{\rm E}} f(\tau_{\rm E}),
 \end{equation}  
which holds for an arbitrary function $f(t)$ which is a slow function of its argument on the time scale $\lambda^{-1}$. In this equation, the Ehrenfest time is defined as
\begin{equation} 
  \tau_{\rm E} = 
  \lambda^{-1} \ln (c^2/\hbar).
\end{equation}
One notes that this definition is consistent with Eq.\ (\ref{eq:tauE}) of the main text, since $c^2$ is a classical action characteristic of the classical motion, which can also be expressed as $c^2 = p_{\rm F} a$, where $a$ is a length scale characteristic of the classical motion. The equivalence to Eq.\ (\ref{eq:tauE}) then follows since $p_{\rm F} = 2 \pi \hbar/\lambda_{\rm F}$. With the equality (\ref{eq:enc1}) the result (\ref{eq:K2a-1}) of the main text follows immediately.

To prove Eq.\ (\ref{eq:enc1}), one makes use of the identity
\begin{equation}
  \frac{2}{\pi} \int_0^{e^{-\lambda t}} dx \frac{\sin x}{x}
  = \theta(-t),
  \label{eq:delta0}
\end{equation}
where the Heaviside step function $\theta(t)$ is broadened on the scale $\lambda^{-1}$. Taking derivatives on each side, we obtain
\begin{eqnarray}
  \label{eq:delta}
  \frac{2\lambda}{\pi} \sin\left(e^{-\lambda t}\right) &=& \delta(t), \\
  \label{eq:deltadt}
  \frac{2\lambda^2}{\pi} e^{-\lambda (t)} \cos\left(e^{-\lambda t}\right) &=& 
  -\frac{\partial}{\partial t}\delta(t).
\end{eqnarray}
These equations can be applied to the left hand side of Eq.\ (\ref{eq:enc1}) after writing the integration in terms of positive $s$ and $u$ and after a variable change that uses $t_{\rm enc}$ and $v=s/c$ as the integration variables,
\begin{eqnarray}
  \lefteqn{\frac{1}{2\pi\hbar}\int_{-c}^{c} ds du \frac{e^{isu/\hbar}}{t_s+t_u} f(t_s+t_u)} \nonumber \\ &=&
  \frac{2}{\pi}\int_{0}^{\infty} dt_{\rm enc} \int_{e^{-\lambda t_{\rm enc}}}^1 dv \frac{\lambda e^{-\lambda (t_{\rm enc}-\tau_{\rm E})}}{v}
  \nonumber \\ && \mbox{} \times \frac{\cos(e^{-\lambda(t_{\rm enc}-\tau_{\rm E})})}{t_{\rm enc}} f(t_{\rm enc})\nonumber\\
  \nonumber \\
  &=&\frac{2\lambda^2}{\pi}\int_{0}^{\infty} dt_{\rm enc} e^{-\lambda (t_{\rm enc}-\tau_{\rm E})}\cos(e^{-\lambda(t_{\rm enc}-\tau_{\rm E})}) f(t_{\rm enc})\nonumber\\
  &=&\frac{\partial}{\partial \tau_{\rm E}} f(\tau_{\rm E}),
\end{eqnarray}
where we used Eq.\ \eqref{eq:deltadt} in the last line.
  
For the contribution $\delta G_{\rm AA}^{{\rm F},2b}$, the encounter is bounded to the right by the phase space point $\vX_2$ of the interaction. In this case, the encounter time is given by $t_{\rm enc} = \lambda^{-1} \ln(c/|s|) + t_{\alpha}$, where $t_{\alpha}$ can take values between zero and $\lambda^{-1} \ln(c/|u|)$. The expression for $\delta G_{\rm AA}^{\rm F,2b}$ is again of the form \eqref{eq:dGAA-F1}, with $K^1$ replaced by
\begin{eqnarray}
 \label{eq:K2b}
  \lefteqn{K^{2b}(\vX_1,\vX_2;\omega)} \nonumber\\
  &=&-\int d\vX P_{\rm in}(\vX) P(\vX,\vX_1;\omega) P_{\rm out}(\vX_2)  \frac{1}{2\pi\hbar}
  \nonumber \\ && \mbox{} \times
  \int_{-c}^c ds du  \int_0^{\lambda^{-1} \ln(c/|u|)} dt_{\alpha} 
  \nonumber \\ && \mbox{} \times
  P(\vX_2,\vX;t_{\rm enc}) e^{i\omega t_{\rm enc}/\hbar}\frac{e^{isu/\hbar}}{t_{\rm enc}}.
\end{eqnarray}
The integration over $s$ and $u$ in Eq.\ \eqref{eq:K2b} is done with the help of the identity
\begin{equation}
  \label{eq:enc2}
  \frac{1}{2\pi\hbar}\int_{-c}^{c} ds du \int_{0}^{\lambda^{-1} \ln(c/|u|)_u}
  dt_{\alpha} \frac{e^{isu/\hbar}}{t_{\rm enc}} f(t_{\rm enc})= f(\tau_{\rm E}),
\end{equation}
which is proven by first performing the integrations over $s$ and $u$, and then using the identity (\ref{eq:delta}). The final result is Eq.\ (\ref{eq:K2b-1}) of the main text. The derivation of Eq.\ (\ref{eq:K2c}) of the main text is similar.

Finally, for the fourth and last contribution $\delta G_{\rm AA}^{\rm F,2d}$, the encounter is bounded by both phase space points $\vX_1$ and $\vX_2$ of the interaction vertices. The encounter time is here given by $t_{\rm enc} = t_{\alpha} + t_{\beta}$, where $t_{\alpha}$ and $t_{\beta}$ vary between $0$ and $\lambda^{-1} \ln(c/|u|)$ and between $0$ and $\lambda^{-1} \ln(c/|s|)$, respectively. The expression for $\delta G_{\rm AA}^{\rm F,2d}$ takes the form \eqref{eq:dGAA-F1}, with $K^1$ replaced by
\begin{eqnarray}
 \label{eq:K2d}
  \lefteqn{K^{2d}(\vX_1,\vX_2;\omega)} \nonumber\\
  &=&-\int d\vX P_{\rm in}(\vX_1)  P_{\rm out}(\vX_2) \frac{1}{2\pi\hbar}  \int_{-c}^c ds du
  \nonumber \\ && \mbox{} \times
  \int_0^{\lambda^{-1} \ln(c/|s|)} dt_{\beta} 
  \int_0^{\lambda^{-1} \ln(c/|u|)} dt_{\alpha} \nonumber\\ 
  && \mbox{} \times   P(\vX_2,\vX;t_{\rm enc}) e^{i\omega(t_{\rm enc})/\hbar}\frac{e^{isu/\hbar}}{t_{\rm enc}}.
\end{eqnarray}
The integrations over $s$ and $u$ can be performed with the help of Eq.\ (\ref{eq:delta0}) and yield Eq.\ (\ref{eq:K2d-1}) of the main text.

\section{Details of the discussion}
\label{sec:appdiscussion}

In this appendix we add some technical details to the discussion of Sec.\ \ref{sec:antidot}. The function $g_m(x)$ of Eq.\ \eqref{eq:gm} can be cast in the following closed analytic expression
\begin{align}
 \label{eq:gmexpl}
 g_m(x)=&\frac{8}{\pi ^2 \left(m^2+x\right)^2}+\left.\frac{32 m^2 a \tanh^s \left(a \frac{\pi  \sqrt{x}}{2}\right)}{\pi ^3 \sqrt{x}
   \left(m^2+x\right)\left(m^2+a x\right)^2}\right|_{a\rightarrow 0}^{a=1}\nonumber\\
    &+\frac{24 F_0^{\sigma} m^2}{\pi ^2 \left(m^2+x\right)^2 \left((1+F_0^{\sigma}) m^2+x\right)}\nonumber\\
   &-\left.\frac{96 b^{3/2} m^2 \tanh^s \left(\frac{\pi  \sqrt{x}}{2
   \sqrt{b}}\right)}{\pi ^3 \sqrt{x} \left(m^2+x\right) \left(b
   m^2+x\right)^2}\right|_{b=1}^{b=1+F_0^{\sigma}},
\end{align}
where $s=\pm 1$ for $m$ even (odd).
In the regime $\tau_{\rm D}\gg\tau_T,\tau_{\rm E}$, we may write this as
\begin{align}
 g_m(n\tau_{\rm D}/\tau_T)=&\frac{8}{\pi^2} \frac{\tau_T^2}{\tau_{\rm D}^2} \frac{1}{(m^2 \tfrac{\tau_T}{\tau_{\rm D}}+n)^2}\nonumber\\
  &\times\left[1+\frac{3 F_0^{\sigma}}{m^2 \tfrac{\tau_T}{\tau_{\rm D}}(1+F_0^{\sigma})+n}\right]\nonumber\\
   &\times\left(1+\mathcal{O}(\sqrt{\tfrac{\tau_T}{\tau_{\rm D}}})\right),
\end{align}
where terms with an additional factor $\sqrt{{\tau_T}/{\tau_{\rm D}}}\propto\frac{1}{L}$ correspond to finite size corrections and will be neglected.

After replacing the summation over $m$ in Eq.\ \eqref{eq:deltaGsum} by an integration, we obtain
\begin{eqnarray}
 \delta G_{\rm AA} &=& -\frac{4 e^2}{\pi h} \sqrt{\frac{\tau_T}{\tau_{\rm D}}}\sum_{n=1}^{\infty} e^{-n\frac{\tau_{\rm E}}{\tau_T}} n \left( \frac{\tau_{\rm E}}{\tau_T} \right)^{5/2} 
  \nonumber \\ && \mbox{} \times
  \left[f_1(n\tfrac{\tau_{\rm E}}{\tau_T})-f'_1(n\tfrac{\tau_{\rm E}}{\tau_T})\right],
\end{eqnarray}
with
\begin{equation}
 f_1(x)= \int_0^{\infty} dz \frac{1}{\sqrt{z}}\frac{1}{(z+x)^2}\left[1+\frac{3F_0^{\sigma}z}{z(1+F_0^{\sigma})+x}\right]e^{-z}.
\end{equation}
Apart from the prefactor $\sqrt{{\tau_T}/{\tau_{\rm D}}}$, the interaction correction $\delta G_{\rm AA}$ is a function of the ratio ${\tau_{\rm E}}/{\tau_T}$ only. The limiting behavior for small and large ratios ${\tau_{\rm E}}/{\tau_T}$ is given in the main text.

For the case $\tau_T\gg\tau_{\rm E},\tau_{\rm D}$, we replace the summation over $n$ in Eq.\ \eqref{eq:deltaGsum} by an integration and find
\begin{eqnarray}
  \delta G_{\rm AA} &=&-\frac{e^2}{h} \sum_{m=1}^{\infty} e^{-m^2 \tfrac{\tau_{\rm E}}{\tau_{\rm D}}} 
  \nonumber \\ && \mbox{} \times
\int_0^{\infty} dx (1+\tfrac{\tau_{\rm E}}{\tau_{\rm D}} x) g_m(x) e^{-x\tfrac{\tau_{\rm E}}{\tau_{\rm D}}}.
 \end{eqnarray}
This is a function of the ratio ${\tau_{\rm E}}/{\tau_{\rm D}}$ only. 

For the case of a large two-dimensional antidot array we use the residue technique to perform the $\omega$ integration of Eq.\ \eqref{eq:dG2d}, and obtain
\begin{eqnarray}
 \delta G_{\rm AA} &=&-\frac{e^2}{h} \frac{W}{L} \frac{1}{\pi} \sum_{n=1}^{\infty}e^{-n\frac{\tau_{\rm E}}{\tau_T}} n\frac{\tau_{\rm E}^2}{\tau_T^2}
  \nonumber \\ && \mbox{} \times
   \left[f_2\left(n\tfrac{\tau_{\rm E}}{\tau_T}\right)-f_2'\left(n\tfrac{\tau_{\rm E}}{\tau_T}\right)\right],
\end{eqnarray}
with
\begin{equation}
 f_2(x)=\int_0^{\infty} dz \frac{1}{(z+x)^2}\left[1+\frac{3F_0^{\sigma}z}{z(1+F_0^{\sigma})+x}\right]e^{-z}.
\end{equation}

\end{document}